\documentclass{article}

\PassOptionsToPackage{numbers}{natbib}
\usepackage[preprint]{neurips_2026}

\usepackage[utf8]{inputenc}
\usepackage[T1]{fontenc}
\usepackage{url}
\usepackage{booktabs}
\usepackage{siunitx}
\usepackage{amsfonts}
\usepackage{microtype}
\usepackage{xcolor}
\usepackage{colortbl}
\definecolor{methodblue}{HTML}{E7F1FE}
\usepackage{pifont}
\usepackage{amsmath}
\usepackage{graphicx}
\usepackage{fontawesome5}
\definecolor{refgreen}{HTML}{4F7F67}
\usepackage[colorlinks=true,
            citecolor=refgreen,
            linkcolor=refgreen,
            urlcolor=refgreen]{hyperref}
\usepackage{multirow}

\usepackage{xspace}
\def\method{PPG2ECG\xspace}

\title{\method{}: Physiology-Guided Latent Rectified Flow for Electro-Hemodynamic PPG-to-ECG Generation}

\author{
\textbf{Xiaoda Wang$^{1,3}$\thanks{Correspondence to: Xiaoda Wang \textless\href{mailto:xiaoda.wang@emory.edu}{\nolinkurl{xiaoda.wang@emory.edu}}\textgreater.}\,,
Minxiao Wang$^{2}$,
Kaiqiao Han$^{3}$,
Defu Cao$^{4}$,
Ching Chang$^{3}$,
Yidan Shi$^{3}$,}\\
\textbf{Runze Yan$^{2}$,
Xiao Luo$^{5}$,
Yan Liu$^{4}$,
Xiao Hu$^{2}$,
Yizhou Sun$^{3}$,
Wei Wang$^{3}$,
Carl Yang$^{1}$}\\[2mm]
\normalfont \small
$^{1}$Department of Computer Science, Emory University\\
\small
$^{2}$Nell Hodgson Woodruff School of Nursing, Emory University\\
\small
$^{3}$Department of Computer Science, University of California, Los Angeles\\
\small
$^{4}$Department of Computer Science, University of Southern California\\
\small
$^{5}$Department of Statistics, University of Wisconsin--Madison\\[1mm]
\normalfont\small \faGithub\ \textbf{Code:} \url{https://github.com/ignite-abd/PPG2ECG}
}

\begin{document}

\maketitle

\begin{abstract}

Electrocardiography (ECG) is the clinical standard for cardiac assessment but requires dedicated hardware that does not scale to daily-life monitoring. Photoplethysmography (PPG) is ubiquitous in wearables but lacks ECG-specific diagnostic morphology and is corrupted by motion and sensor noise. PPG-to-ECG generation aims to bridge this gap by recovering electrical morphology and timing from peripheral pulse signals.
However, existing methods largely rely on statistical alignment and data-driven generation. They fail to explicitly structure the latent space around physiology-aware electro-hemodynamic factors and lack constraints from forward physiological dynamics.
To address these challenges, we propose \textbf{\method{}}, a physiology-guided latent rectified flow framework. \method{} introduces an electro-hemodynamic simulator that co-models ECG and PPG through shared cardiac phase dynamics. Guided by this simulator, a Physiology-Aware AutoEncoder learns a structured electro-hemodynamic latent space. Then we integrate this simulator guidance into a PPG-conditioned latent rectified flow, enforcing ECG-side morphology consistency and ECG-to-PPG forward hemodynamic consistency during generative transport.
Experiments on the large-scale MC-MED dataset demonstrate that \method{} significantly improves PPG-to-ECG generation and downstream cardiovascular disease classification, proving its ability to generate ECGs that are both signal-faithful and physiologically plausible under the ECG-to-PPG hemodynamic pathway.

\end{abstract}

\section{Introduction}

Electrocardiography (ECG) is the clinical standard for measuring cardiac electrical activity and diagnosing cardiovascular conditions~\citep{kligfield2007recommendations,wang2026position,berkaya2018survey,wang2026beyond,jin2025reading,jin2026ecg}.
However, continuous ECG monitoring requires dedicated electrodes and stable skin contact, limiting its scalability in daily-life settings~\citep{wang2025unified}. Photoplethysmography (PPG), in contrast, is widely available in consumer wearables, offering a low-cost measurement of peripheral blood-volume changes~\citep{allen2007photoplethysmography,wang2026ppg,charlton2022wearable,zhang2026deep,nie2026anyppg,xu2026inertia}. Despite its appeal, PPG is fundamentally limited by high levels of motion artifacts and the inherent absence of diagnostic electrical features like P-QRS-T morphology~\citep{fine2021sources,castaneda2018review,chen2025gpt}. As a delayed hemodynamic response, raw PPG is structurally deficient for direct and reliable clinical diagnosis. Overcoming this bottleneck by generating diagnostically informative ECGs from wearable PPG therefore provides a highly meaningful bridge between ubiquitous sensing and verifiable cardiac assessment~\citep{sarkar2021cardiogan,tang2023ppg2ecgps,shome2024region,wang2026se}.

Recent PPG-to-ECG studies have progressed from supervised waveform reconstruction to adversarial, diffusion-based, token-based, and latent flow-based generation~\citep{tang2023ppg2ecgps,sarkar2021cardiogan,shome2024region,ding2025ai,fang2025ppgflowecg}. Despite these advances, most existing methods treat PPG-to-ECG generation primarily as a data-driven cross-modal translation problem, leaving two key challenges:
(i) \textit{Lack of physiology-aware latent abstraction.}
PPG and ECG are coupled but asymmetric observations of the same cardiovascular process: ECG reflects upstream electrical activation, whereas PPG captures the delayed peripheral pulse~\citep{bao2026rethinking}. Purely statistical alignment fails to explicitly encode this directional physiological coupling.
(ii) \textit{Missing physiological simulator guidance.}
Existing methods optimize data-driven signal similarity and easily hallucinate physically impossible hemodynamic delays, ignoring cardiac-cycle dynamics and the forward ECG-to-PPG hemodynamic pathway~\citep{tang2023ppg2ecgps}. Incorporating this knowledge is non-trivial; naively imposing a fixed simulator as a hard constraint risks severe model mismatch and over-regularization. A robust approach must instead integrate simulator knowledge as soft, phase-aware guidance while retaining the flexibility of data-driven latent generation.

To address these challenges, we propose \method{}, a physiology-guided latent rectified flow framework for electro-hemodynamic PPG-to-ECG generation.
We introduce a novel electro-hemodynamic simulator that explicitly co-models ECG and PPG by linking an upstream electrical readout with a delayed peripheral pulse via shared cardiac phase dynamics. Using this simulator as a physiological anchor, \method{} employs a Physiology-Aware AutoEncoder (PAE) to abstract modality-specific observations into a structured cardiovascular latent space regularized by cross-modal alignment and latency-aware consistency. On top of this latent space, \method{} introduces a simulator-guided latent rectified flow. During training, the generative transport is explicitly regularized by the simulator to ensure both ECG-side morphological fidelity and forward ECG-to-PPG hemodynamic consistency.
In summary, our main contributions are as follows:

\ding{182} \textbf{\textit{Physiology-Guided Generative Formulation.}}
We introduce \method{}, formulating PPG-to-ECG generation as an electro-hemodynamic generation problem. We design an electro-hemodynamic physiological simulator to co-model ECG and PPG through shared cardiac phase dynamics, upstream electrical readouts, and delayed hemodynamic readouts.

\ding{183} \textbf{\textit{Physiology-Aware Representation Learning.}}
We design a Physiology-Aware AutoEncoder that abstracts modality-specific observations into a structured electro-hemodynamic latent space, rigorously constrained by simulator-derived latency-aware consistency, and cross-modal latent alignment.

\ding{184} \textbf{\textit{Simulator-Guided Flow Matching.}}
We integrate the electro-hemodynamic simulator into PPG-conditioned latent rectified flow. Simulator-derived constraints guide the model to generate ECGs that are waveform-faithful and physiologically plausible under the ECG-to-PPG hemodynamic pathway.

\ding{185} \textbf{\textit{Multi-Level Experimental Validation.}}
We demonstrate that \method{} consistently improves PPG-to-ECG generation across waveform fidelity, clinical interval consistency, and cardiovascular disease detection on the large-scale MC-MED dataset.

\section{Related Work}
\label{sec:related_work}

\noindent \textbf{PPG-to-ECG Generation.}
PPG-to-ECG generation aims to recover cardiac electrical waveforms from PPG. One line of work treated this as supervised signal reconstruction, utilizing CNNs, RNNs, or Transformers~\citep{lan2023performer} and specialized architectures like W-Net~\citep{tang2023ppg2ecgps}. To enhance waveform realism, another line of work adopted generative models. For example, CardioGAN~\citep{sarkar2021cardiogan} and CLEP-GAN~\citep{li2025clep} employed adversarial and contrastive learning to improve subject-independent translation.
More recently, diffusion and flow-based models have driven state-of-the-art high-fidelity translation. RDDM~\citep{shome2024region} utilizes a region-disentangled diffusion process for targeted ECG components, while CardioPPG~\citep{ding2025ai} generates ECGs through a tokenized frequency-domain pathway. Most closely related to our work, PPGFlowECG~\citep{fang2025ppgflowecg} employs a two-stage latent rectified flow, generating ECGs from a shared cross-modal latent space. Despite these advancements, existing models rely predominantly on statistical data-driven alignment, lacking explicit regularization from forward physiological dynamics.

\noindent \textbf{Physiological Simulators.}
Compact mathematical models utilize low-dimensional differential equations to simulate the cardiac cycle and generate stereotyped cardiovascular waveforms~\citep{mcsharry2003dynamical, edelmann2018ecg}. The canonical ECGSYN model~\citep{mcsharry2003dynamical}, for instance, employs a three-dimensional limit-cycle oscillator with phase-locked Gaussian components to synthesize realistic P--QRS--T morphologies and heart rate variability. Similar oscillator-based frameworks have since been developed to model delayed peripheral hemodynamics, such as PPG~\citep{tang2020synthetic}. However, purely mechanistic simulators rely on globally fixed templates, limiting their expressivity in capturing subject-specific conduction abnormalities or complex cross-modal variations. To overcome this, recent hybrid approaches integrate physiological priors with Neural ODEs~\citep{chen2018neural,wang2025conditional,liu2025graph} or Universal Differential Equations~\citep{rackauckas2021universal}. They can enable models to preserve essential physiological structures while flexibly fitting real-world clinical data~\citep{golany2020simgans,golany2021ecgodegan,wang2026se}.

\noindent \textbf{Latent Generative Models.}
Diffusion and score-based generative models have become powerful tools for learning complex data distributions by gradually transforming noise into structured samples~\citep{ho2020denoising,croitoru2023diffusion,yang2023diffusion,qian2025physdiff,li2025eddm}. To improve computational efficiency, latent diffusion models move the generative process from the raw data space to a compressed latent space, reducing the cost of modeling high-dimensional signals while preserving sample quality~\citep{rombach2022high}. This idea is particularly relevant for physiological waveform generation, where long, high-frequency, and noisy sensor sequences make direct generative modeling challenging~\citep{shuai2026signal}.
Flow-based generative models provide an alternative continuous-time view of generation. Flow matching learns vector fields that transport simple base distributions to data distributions without simulating stochastic diffusion during training~\citep{lipman2023flow,stoica2025contrastive,wildberger2023flow}. Rectified flow further encourages trajectories to follow straight transport paths between source and target distributions, enabling efficient sampling with coarse numerical integration~\citep{liu2023flow,lee2024improving,esser2024scaling}. These properties make rectified flow attractive for PPG-to-ECG generation, where the model must transport PPG-conditioned latent variables toward plausible ECG representations. Existing latent flow methods for PPG-to-ECG generation have shown strong empirical performance~\citep{fang2025ppgflowecg}, but their transport objectives are still largely statistical.

\noindent \textbf{Cross-Modal Physiological Representation Learning.}
Learning aligned representations across physiological modalities is central to translating PPG into ECG. Since PPG and ECG are generated by different but coupled physiological processes, successful cross-modal learning must balance modality-specific feature extraction with shared cardiovascular abstraction. Prior methods often encourage alignment by using shared encoders, contrastive objectives, or paired latent reconstruction~\citep{nie2026anyppg,chen2026bridging,fang2026echobridge,xu2026sleeplm,shuai2026osf,he2026cap}. CardioPPG aligns PPG and ECG embeddings through cross-modal contrastive learning before using the aligned PPG encoder for downstream cardiovascular prediction and ECG synthesis~\citep{ding2025ai}. PPGFlowECG similarly builds a shared latent space through a CardioAlign encoder and performs ECG generation in that aligned space~\citep{fang2025ppgflowecg}. These approaches, together with latent cross-modal synthesis in other clinical modalities~\citep{fang2026ecgflowcmr,wu2026mirage}, demonstrate that latent alignment can improve downstream utility and generation quality.

\section{Method: \method{}}
\label{sec:method}

\begin{figure}[t]
    \centering
    \includegraphics[width=1.0\linewidth]{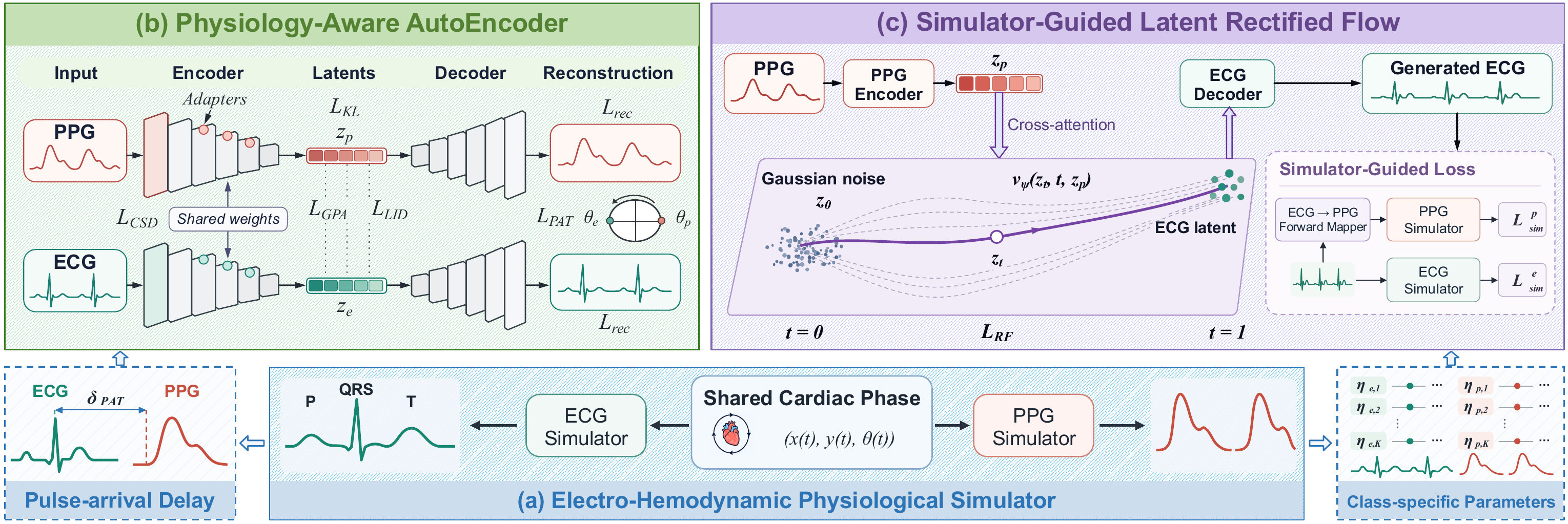}
\caption{\textbf{Overview Framework of \method{}.}
(a) \textit{The Electro-Hemodynamic Physiological Simulator}: co-models ECG and PPG through shared cardiac phase dynamics.
(b) \textit{Physiology-Aware AutoEncoder}: learns an aligned electro-hemodynamic latent space from resampled PPG and ECG windows.
(c) \textit{Simulator-Guided Latent Rectified Flow}: generates ECG latents conditioned on PPG and regularizes the transport with ECG morphology and ECG-to-PPG hemodynamic consistency.
}
    \label{fig:overview}
\end{figure}

We present \method{}, a physiology-guided latent rectified flow framework for electro-hemodynamic PPG-to-ECG generation. \method{} has three components. First, an electro-hemodynamic physiological simulator co-models ECG and PPG through shared cardiac phase dynamics, an upstream ECG readout, and a delayed PPG readout. Second, a Physiology-Aware AutoEncoder (PAE) learns a modality-aware shared latent space regularized by cross-modal alignment and simulator-derived latency consistency. Third, a PPG-conditioned latent rectified flow generates ECG latents from Gaussian noise, while simulator-derived residuals guide the transport toward ECG-side morphology consistency and ECG-to-PPG forward hemodynamic consistency. The simulator and the ECG-to-PPG forward mapper are used only during training; inference requires only the input PPG.

\noindent\textbf{Problem Formulation.}
Given a paired PPG segment $\mathbf{x}_{p}\in\mathbb{R}^{L_p\times 1}$ and ECG segment $\mathbf{x}_{e}\in\mathbb{R}^{L_e\times 1}$, where $L_p$ and $L_e$ denote the numbers of PPG and ECG samples after resampling, our goal is to learn a conditional generator $p(\mathbf{x}_{e}\mid \mathbf{x}_{p})$ that synthesizes a plausible ECG waveform consistent with the observed PPG. The Physiology-Aware AutoEncoder maps both modalities into aligned latent representations $\mathbf{z}_{p},\mathbf{z}_{e}\in\mathbb{R}^{C_z\times T_z}$, where $C_z$ and $T_z$ denote the latent channel dimension and temporal length. The Electro-Hemodynamic Simulator Guided Latent Rectified Flow then learns a conditional vector field $\mathbf{v}_{\psi}(\mathbf{z}_{t},t,\mathbf{z}_{p})$ that transports a Gaussian latent variable toward the ECG latent distribution conditioned on $\mathbf{z}_{p}$. At inference time, only $\mathbf{x}_{p}$ is required: we encode it into $\mathbf{z}_{p}$, solve the learned flow to obtain $\hat{\mathbf{z}}_{e}$, and decode it into the generated ECG $\hat{\mathbf{x}}_{e}$.

\subsection{The Electro-Hemodynamic Physiological Simulator}
\label{subsec:eh_simulator}

Because PPG is a delayed hemodynamic response rather than a direct electrical measurement, effective PPG-to-ECG generation requires a physiological anchor that captures the asymmetric ECG-to-PPG coupling. We therefore introduce an electro-hemodynamic physiological simulator composed of four ODE functions. The functions $f_x$ and $f_y$ define a shared cardiac phase oscillator, $f_e$ generates the ECG voltage readout $e(t)$, and $f_p$ generates the PPG pulse readout $p(t)$. The oscillator state $(x,y)$ is not a measured signal; it acts as a cardiac clock whose polar angle $\theta$ is the phase shared by both readouts, so that ECG and PPG events are locked to the same beat. This simulator is not used as a rigid waveform template at inference time. Instead, it provides a soft training-time interface: the pulse-arrival delay $\delta_{\mathrm{PAT}}$ regularizes the PAE latent space, while the vector fields $f_e$ and $f_p$ define finite-difference residuals for simulator-guided latent flow training.
Following the ECGSYN dynamical model~\citep{mcsharry2003dynamical} and its use as a simulator prior for ECG generation~\citep{wang2026se}, the shared cardiac phase follows a limit-cycle oscillator:
\begin{equation}
\label{eq:shared_phase}
\begin{aligned}
\frac{dx}{dt} &= f_x(x,y)=\alpha(x,y)x-\omega y, \qquad \frac{dy}{dt} &= f_y(x,y)=\alpha(x,y)y+\omega x,
\end{aligned}
\end{equation}
where $\alpha(x,y)=1-\sqrt{x^2+y^2}$ drives $(x,y)$ toward the unit limit cycle, and $\theta(x,y)=\mathrm{atan2}(y,x)\in(-\pi,\pi]$ denotes the cardiac phase. In polar coordinates $(r,\theta)$, the radius obeys $\dot r=(1-r)r$ while the phase advances at the constant rate $\dot\theta=\omega$. Hence one revolution of $(x,y)$, with period $2\pi/\omega$, corresponds to one cardiac cycle, and $\omega$ sets the average heart rate.

The ECG readout follows the phase-locked Gaussian P--QRS--T formulation of ECGSYN~\citep{mcsharry2003dynamical}:
\begin{equation}
\label{eq:e_readout}
\frac{de}{dt}
=
f_e(x,y,e,t;\eta_e)
=
-\sum_{\beta\in\mathcal{B}_e}
a^e_\beta \Delta\theta^e_\beta(x,y)
\exp\left(
-\frac{(\Delta\theta^e_\beta(x,y))^2}{2(b^e_\beta)^2}
\right)
-
\left[e-e_0(t)\right],
\end{equation}
where $\mathcal{B}_e=\{P,Q,R,S,T\}$,
$\Delta\theta^e_\beta(x,y)=\left(\theta(x,y)-\theta^e_\beta\right)\bmod 2\pi$, and $e_0(t)$ is a slowly varying baseline (e.g., respiratory drift) toward which the readout relaxes.
The parameters $\eta_e$ collect the phase locations, amplitudes, and temporal widths of the ECG components.
Analogous to Gaussian-kernel PPG pulse models~\citep{tang2020synthetic}, the PPG readout is driven by the same cardiac phase but delayed relative to the ECG R-peak:
\begin{equation}
\label{eq:p_readout}
\frac{dp}{dt}
=
f_p(x,y,p,t;\eta_p)
=
-\sum_{\gamma\in\mathcal{B}_p}
a^p_\gamma \Delta\theta^p_\gamma(x,y)
\exp\left(
-\frac{(\Delta\theta^p_\gamma(x,y))^2}{2(b^p_\gamma)^2}
\right)
-
\nu_p\left[p-p_0(t)\right],
\end{equation}
where $\mathcal{B}_p=\{\mathrm{foot},\mathrm{sys},\mathrm{notch},\mathrm{dia}\}$ denotes the p-foot, systolic peak, dicrotic notch, and diastolic wave, $p_0(t)$ is the PPG baseline, and $\nu_p>0$ is the vascular relaxation rate toward it. The phase offset of each PPG component is defined as
\begin{equation}
\label{eq:ppg_delay_phase}
\Delta\theta^p_\gamma(x,y)
=
\left(
\theta(x,y)-\theta^e_R-\delta_{\mathrm{PAT}}-\theta^p_\gamma
\right)\bmod 2\pi,
\end{equation}
where $\theta^e_R$ is the ECG R-peak phase and $\delta_{\mathrm{PAT}}$ denotes the pulse-arrival delay from electrical activation to the peripheral pulse response. The parameters $\eta_p$ collect $\delta_{\mathrm{PAT}}$, the relaxation coefficient $\nu_p$, and the phase, amplitude, and width parameters of the PPG components.

\noindent\textbf{Euler Residual Interface.}
We discretize the simulator on modality-specific time grids using explicit Euler updates.
For modality $m\in\{e,p\}$, let $\tilde{\mathbf{h}}_{m}$ denote a generated beat crop, let
$\mathbf{s}_{m,\ell}=(x_{m,\ell},y_{m,\ell})$ denote the corresponding reference cardiac phase state, and let $f_m$ be the simulator vector field with fitted parameters $\eta_m$.
The simulator induces the finite-difference residual
\begin{equation}
\label{eq:euler_residual_interface}
\mathbf{r}^{m}_{\ell}
=
\frac{
\tilde{h}_{m,\ell+1}
-
\tilde{h}_{m,\ell}
}{
\Delta t_m
}
-
f_m(
\mathbf{s}_{m,\ell},
\tilde{h}_{m,\ell},
t_{m,\ell};
\eta_m
).
\end{equation}
Rather than enforcing the simulator as a hard waveform template, \method{} uses this residual as soft physiological guidance during training.
The full Euler discretization and simulator fitting procedure are provided in Appendix~\ref{app:euler_discretization}.

\subsection{Physiology-Aware AutoEncoder}
\label{subsec:pae}

The Physiology-Aware AutoEncoder (PAE) learns a shared cardiovascular latent space for paired PPG and ECG signals while preserving modality-specific acquisition characteristics. As shown in Figure~\ref{fig:overview}, the encoder adopts a modality-specific-to-shared design. PPG and ECG are first processed by modality-specific blocks that preserve acquisition-dependent characteristics, such as waveform morphology, frequency content, and noise patterns. The resulting representations are then passed through shared blocks to extract common cardiovascular factors. To avoid losing useful modality-dependent information during shared abstraction, we insert lightweight modality-aware adapters in the intermediate shared blocks. A final shared bottleneck maps both modalities to a common latent posterior space.
For modality $m\in\{p,e\}$, let $\mathbf{h}_{m}^{(0)}=\mathbf{x}_{m}$. The modality-specific blocks are defined as
\begin{equation}
\mathbf{h}_{m}^{(i)}
=
F_{m,i}^{\mathrm{spec}}(\mathbf{h}_{m}^{(i-1)}),
\qquad i=1,2,
\end{equation}
where $F_{m,i}^{\mathrm{spec}}$ denotes the $i$-th modality-specific block for modality $m$. The intermediate shared blocks with modality-aware adapters are then written as
\begin{equation}
\mathbf{u}_{m}^{(k)}
=
F_{k}^{\mathrm{sh}}(\mathbf{h}_{m}^{(k-1)}),
\qquad
\mathbf{h}_{m}^{(k)}
=
\mathbf{u}_{m}^{(k)}
+
A_{m,k}(\mathbf{u}_{m}^{(k)}),
\qquad k=3,\ldots,5,
\end{equation}
where $F_{k}^{\mathrm{sh}}$ is the $k$-th shared block and $A_{m,k}$ is the modality-aware adapter. The final shared bottleneck is
\begin{equation}
\mathbf{h}_{m}^{(6)}
=
F_{6}^{\mathrm{sh}}(\mathbf{h}_{m}^{(5)}).
\end{equation}
The encoder parameterizes a diagonal Gaussian posterior by
\begin{equation}
[\boldsymbol{\mu}_{m},\log\boldsymbol{\sigma}_{m}^{2}]
=
H_{\phi}(\mathbf{h}_{m}^{(6)}),
\qquad
q_{\phi}^{m}(\mathbf{z}_{m}\mid \mathbf{x}_{m})
=
\mathcal{N}\left(
\boldsymbol{\mu}_{m},
\mathrm{diag}(\boldsymbol{\sigma}_{m}^{2})
\right),
\end{equation}
and samples the latent representation as
\begin{equation}
\mathbf{z}_{m}
=
\boldsymbol{\mu}_{m}
+
\boldsymbol{\sigma}_{m}\odot\boldsymbol{\epsilon},
\qquad
\boldsymbol{\epsilon}\sim\mathcal{N}(\mathbf{0},\mathbf{I}).
\end{equation}
Here $\boldsymbol{\mu}_{m},\log\boldsymbol{\sigma}_{m}^{2}\in\mathbb{R}^{C_z\times T_z}$, so both PPG and ECG are mapped into aligned latent representations $\mathbf{z}_{p},\mathbf{z}_{e}\in\mathbb{R}^{C_z\times T_z}$.
The decoder for each modality reconstructs the waveform with the same number of samples as its input.
It contains a residual reconstruction branch and a lightweight decomposition branch:
\begin{equation}
\hat{\mathbf{x}}_{m}
=
D_{\theta}^{m}(\mathbf{z}_{m})
=
D_{\mathrm{res}}^{m}(\mathbf{z}_{m})
+
D_{\mathrm{dec}}^{m}(\mathbf{z}_{m}),
\end{equation}
where
\begin{equation}
D_{\mathrm{dec}}^{m}(\mathbf{z}_{m})
=
D_{\mathrm{level}}^{m}(\mathbf{z}_{m})
+
D_{\mathrm{trend}}^{m}(\mathbf{z}_{m})
+
D_{\mathrm{season}}^{m}(\mathbf{z}_{m}).
\end{equation}
The residual branch models local waveform morphology, while the decomposition branch provides a lightweight bias for global level, slow temporal drift, and coarse recurrent structure.

\noindent\textbf{PAE Training Objective.}
PAE is optimized with three groups of objectives: waveform reconstruction at each modality's input rate, cross-modal latent alignment, and simulator-informed delay consistency.

\noindent (i) \textit{Simulator-informed phase delay consistency.}
PPG is a delayed hemodynamic response to ECG activation. Therefore, the phase gap between the generated ECG and PPG should be consistent with the class-specific simulator delay $\delta_{\mathrm{PAT}}$. Let
\(
\hat{\mathbf{x}}_{e}=D_{\theta}^{e}(\mathbf{z}_{e}),
\)
and
\(
\hat{\mathbf{x}}_{p}=D_{\theta}^{p}(\mathbf{z}_{p}),
\)
where $\hat{\mathbf{x}}_{e}$ and $\hat{\mathbf{x}}_{p}$ are the reconstructed ECG and PPG waveforms. Let $\Theta_{e}(\cdot)$ estimate the ECG activation phase and $\Theta_{p}(\cdot)$ estimate the PPG pulse phase. The estimated electro-hemodynamic phase delay is
\(
\widehat{\Delta\theta}_{\mathrm{PAT}}
=
\mathrm{wrap}_{\pi}
\left(
\Theta_{p}(\hat{\mathbf{x}}_{p})
-
\Theta_{e}(\hat{\mathbf{x}}_{e})
\right),
\)
where $\mathrm{wrap}_{\pi}(\cdot)$ maps the phase difference to $[-\pi,\pi]$. Since phase is circular, we use a circular squared distance:
\(
d_{\mathrm{circ}}(a,b)
=
\left[
\mathrm{atan2}
\left(
\sin(a-b),
\cos(a-b)
\right)
\right]^2 .
\)
The phase delay consistency loss is then
\begin{equation}
\mathcal{L}_{\mathrm{PAT}}
=
\frac{1}{B}
\sum_{i=1}^{B}
d_{\mathrm{circ}}
\left(
\widehat{\Delta\theta}_{\mathrm{PAT}}^{(i)},
\delta_{\mathrm{PAT}}^{(i)}
\right).
\end{equation}

\noindent (ii) \textit{Autoencoding regularization.}
The reconstruction loss preserves modality-specific waveform information:
\begin{equation}
\mathcal{L}_{\mathrm{rec}}
=
\left\|
D_{\theta}^{p}(\mathbf{z}_{p})-\mathbf{x}_{p}
\right\|_{2}^{2}
+
\left\|
D_{\theta}^{e}(\mathbf{z}_{e})-\mathbf{x}_{e}
\right\|_{2}^{2}.
\end{equation}
The KL term regularizes both modality posteriors toward a standard Gaussian prior:
\begin{equation}
\mathcal{L}_{\mathrm{KL}}
=
D_{\mathrm{KL}}
\left(
q_{\phi}^{p}(\mathbf{z}_{p}\mid\mathbf{x}_{p})
\,\|\,
\mathcal{N}(\mathbf{0},\mathbf{I})
\right)
+
D_{\mathrm{KL}}
\left(
q_{\phi}^{e}(\mathbf{z}_{e}\mid\mathbf{x}_{e})
\,\|\,
\mathcal{N}(\mathbf{0},\mathbf{I})
\right).
\end{equation}

\noindent (iii) \textit{Cross-modal latent alignment.}
We align paired PPG and ECG representations at both distribution and instance levels. Global posterior alignment matches posterior means and dispersion:
\begin{equation}
\mathcal{L}_{\mathrm{GPA}}
=
\left\|
\boldsymbol{\mu}_{e}
-
\boldsymbol{\mu}_{p}
\right\|_{2}^{2}
+
\frac{1}{2}
\left[
D_{\mathrm{KL}}
\left(
q_{\phi}^{e}(\mathbf{z}_{e}\mid\mathbf{x}_{e})
\,\|\,
q_{\phi}^{p}(\mathbf{z}_{p}\mid\mathbf{x}_{p})
\right)
+
D_{\mathrm{KL}}
\left(
q_{\phi}^{p}(\mathbf{z}_{p}\mid\mathbf{x}_{p})
\,\|\,
q_{\phi}^{e}(\mathbf{z}_{e}\mid\mathbf{x}_{e})
\right)
\right].
\end{equation}
To preserve instance-level distinctions, we use a bidirectional contrastive loss. Let $\bar{\mathbf{z}}_{m}^{(i)}$ denote the normalized pooled latent of the $i$-th sample from modality $m$, and define $s(\mathbf{a},\mathbf{b})=\exp(\langle\mathbf{a},\mathbf{b}\rangle/\tau)$.
\begin{equation}
\mathcal{L}_{\mathrm{LID}}
=
-\frac{1}{2B}
\sum_{i=1}^{B}
\left[
\log
\frac{
s(\bar{\mathbf{z}}_{p}^{(i)},\bar{\mathbf{z}}_{e}^{(i)})
}{
\sum_{j=1}^{B}
s(\bar{\mathbf{z}}_{p}^{(i)},\bar{\mathbf{z}}_{e}^{(j)})
}
+
\log
\frac{
s(\bar{\mathbf{z}}_{e}^{(i)},\bar{\mathbf{z}}_{p}^{(i)})
}{
\sum_{j=1}^{B}
s(\bar{\mathbf{z}}_{e}^{(i)},\bar{\mathbf{z}}_{p}^{(j)})
}
\right].
\end{equation}
Finally, we use a weak cross-modal decodability objective to encourage each latent to retain modality-shared physiological information. And this objective is assigned a small weight and is intended only as a weak physiological decodability constraint.
\begin{equation}
\mathcal{L}_{\mathrm{CSD}}
=
\left\|
D_{\theta}^{e}(\mathbf{z}_{p})-\mathbf{x}_{e}
\right\|_{2}^{2}
+
\left\|
D_{\theta}^{p}(\mathbf{z}_{e})-\mathbf{x}_{p}
\right\|_{2}^{2}.
\end{equation}

The full PAE objective is
\begin{equation}
\mathcal{L}_{\mathrm{PAE}}
=
\lambda_{\mathrm{PAT}}\mathcal{L}_{\mathrm{PAT}}
+
\mathcal{L}_{\mathrm{rec}}
+
\beta\mathcal{L}_{\mathrm{KL}}
+
\lambda_{\mathrm{GPA}}\mathcal{L}_{\mathrm{GPA}}
+
\lambda_{\mathrm{LID}}\mathcal{L}_{\mathrm{LID}}
+
\lambda_{\mathrm{CSD}}\mathcal{L}_{\mathrm{CSD}}.
\end{equation}
After PAE training, the PPG encoder, ECG encoder, and ECG decoder are frozen for the subsequent PPG-conditioned latent rectified flow.

\subsection{Electro-Hemodynamic Simulator Guided Latent Rectified Flow}
\label{subsec:sim_guided_lrf}

After PAE training, we freeze the PPG encoder, ECG encoder, and ECG decoder. For each paired sample, we obtain the PPG condition latent $\mathbf{z}_{p}=E_{\phi}^{p}(\mathbf{x}_{p})$ and the target ECG latent $\mathbf{z}_{e}=E_{\phi}^{e}(\mathbf{x}_{e})$. Following rectified flow~\citep{liu2023flow} and its latent PPG-to-ECG variant~\citep{fang2025ppgflowecg}, we learn a conditional vector field that transports Gaussian noise toward the ECG latent distribution conditioned on $\mathbf{z}_{p}$. Given $\mathbf{z}_{0}\sim\mathcal{N}(\mathbf{0},\mathbf{I})$ and $t\sim\mathcal{U}(0,1)$, we define the linear interpolation path
\begin{equation}
\label{eq:rf_path}
\mathbf{z}_{t}
=
(1-t)\mathbf{z}_{0}
+
t\mathbf{z}_{e},
\end{equation}
and optimize the flow-matching objective
\begin{equation}
\label{eq:rf_loss}
\mathcal{L}_{\mathrm{RF}}
=
\mathbb{E}
\left[
\left\|
\mathbf{v}_{\psi}(\mathbf{z}_{t},t,\mathbf{z}_{p})
-
(\mathbf{z}_{e}-\mathbf{z}_{0})
\right\|_{2}^{2}
\right].
\end{equation}
The vector field $\mathbf{v}_{\psi}$ is parameterized by a conditional Transformer, where the evolving ECG latent attends to the PPG condition latent through cross-attention.
To further constrain the generated ECG with electro-hemodynamic dynamics, we apply simulator guidance only to the terminal generated state rather than to every intermediate flow state. This avoids unstable gradients from early noisy latent states and aligns the physiological regularization with the final ECG waveform used at inference.
Given a noise latent $z_0$, we obtain a differentiable terminal estimate by integrating the learned vector field with $K$ Euler steps:
\begin{equation}
\label{eq:terminal_state_estimate}
\bar{z}_{k+1}
=
\bar{z}_k
+
\Delta \tau\,
v_\psi(\bar{z}_k,\tau_k,z_p),
\quad
\tau_k=\frac{k}{K},
\quad
\Delta\tau=\frac{1}{K},
\qquad
\tilde{z}_e=\bar{z}_K,\quad
\tilde{x}_e=D^e_\theta(\tilde{z}_e).
\end{equation}
The decoded terminal ECG $\tilde{x}_e$ is then used to compute simulator-informed residual losses.

\noindent\textbf{ECG-to-PPG Forward Mapper.}
We independently train an ECG-to-PPG conditional latent rectified flow $G_\rho$ on paired ECG--PPG windows from the MC-MED training split, with the PAE frozen. Its flow-matching objective conditions on the ECG latent and targets the paired PPG latent. During \method{} training, the decoded terminal ECG $\tilde{x}_e$ induces the pulse waveform used in the PPG-side simulator residual,
\begin{equation}
\label{eq:ecg_to_ppg_mapper}
\tilde{x}_{p}=G_\rho(\tilde{x}_{e}).
\end{equation}
We integrate the mapper's vector field with 10 Euler steps and decode the resulting PPG latent with the frozen PAE. The mapper's parameters remain fixed while the PPG-side residual backpropagates through $G_\rho$ to the ECG generator. Inference from PPG omits $G_\rho$.

\noindent\textbf{Simulator-guided Residual Losses.}
To capture heterogeneous electro-hemodynamic patterns, we fit simulator parameters using ICD-stratified training groups. For each group $\kappa$, we estimate $\eta^\kappa=(\eta^\kappa_e,\eta^\kappa_p)$ from representative ECG--PPG beats. These parameters are used only during training: they define the simulator residuals below, and their delay $\delta_{\mathrm{PAT}}$ also provides the target of $\mathcal{L}_{\mathrm{PAT}}$ in PAE training. ICD labels are not used as model inputs, and \method{} requires only PPG at inference time.
Let $\mathcal{C}_{e}(\tilde{\mathbf{x}}_{e}^{t})=\tilde{\mathbf{h}}_{e}\in\mathbb{R}^{L_{e}^{c}}$ denote a QRS-aligned generated ECG beat crop. Given fitted ECG simulator parameters $\eta_{e}^{\kappa}$ and reference phase states $(x_{e,\ell}^{\kappa},y_{e,\ell}^{\kappa})$, we penalize deviations from the ECG vector field by the Euler residual:
\begin{equation}
\label{eq:ecg_sim_loss}
\mathcal{L}_{\mathrm{sim}}^{e}
=
\frac{1}{L_{e}^{c}-1}
\sum_{\ell=1}^{L_{e}^{c}-1}
\left\|
\frac{
\tilde{h}_{e,\ell+1}
-
\tilde{h}_{e,\ell}
}{
\Delta t_{e}
}
-
f_{e}
\left(
x_{e,\ell}^{\kappa},
y_{e,\ell}^{\kappa},
\tilde{h}_{e,\ell},
t_{e,\ell};
\eta_{e}^{\kappa}
\right)
\right\|_{2}^{2}.
\end{equation}
This term encourages generated ECG beats to follow physiologically plausible P--QRS--T dynamics rather than only matching point-wise waveform statistics.

Similarly, let $\mathcal{C}_{p}(\tilde{\mathbf{x}}_{p}^{t})=\tilde{\mathbf{h}}_{p}\in\mathbb{R}^{L_{p}^{c}}$ denote the induced PPG beat crop. Given fitted PPG simulator parameters $\eta_{p}^{\kappa}$ and reference phase states $(x_{p,\ell}^{\kappa},y_{p,\ell}^{\kappa})$, we define
\begin{equation}
\label{eq:ppg_sim_loss}
\mathcal{L}_{\mathrm{sim}}^{p}
=
\frac{1}{L_{p}^{c}-1}
\sum_{\ell=1}^{L_{p}^{c}-1}
\left\|
\frac{
\tilde{h}_{p,\ell+1}
-
\tilde{h}_{p,\ell}
}{
\Delta t_{p}
}
-
f_{p}
\left(
x_{p,\ell}^{\kappa},
y_{p,\ell}^{\kappa},
\tilde{h}_{p,\ell},
t_{p,\ell};
\eta_{p}^{\kappa}
\right)
\right\|_{2}^{2}.
\end{equation}
This term regularizes the generated ECG through its induced PPG response, encouraging compatibility with delayed peripheral pulse dynamics.

The complete simulator-guided flow objective is
\begin{equation}
\label{eq:flow_objective}
\mathcal{L}_{\mathrm{Flow}}
=
\mathcal{L}_{\mathrm{RF}}
+
\lambda_{e}\mathcal{L}_{\mathrm{sim}}^{e}
+
\lambda_{p}\mathcal{L}_{\mathrm{sim}}^{p}.
\end{equation}
The simulator-informed losses and the ECG-to-PPG forward mapper are used only during training. At inference time, the model only requires the PPG input, solves the learned latent flow, and decodes the generated ECG latent through the frozen ECG decoder.

\subsection{Training and Inference}
\label{subsec:training_objective}

\method{} is optimized in two phases. We first train the Physiology-Aware AutoEncoder with the objective $\mathcal{L}_{\mathrm{PAE}}$ to obtain aligned PPG and ECG latent representations. Afterward, we freeze the PPG encoder, ECG encoder, and ECG decoder, and train only the PPG-conditioned latent rectified flow.
The simulator-informed terms are used only during training and do not change the inference procedure.
At inference time, only the PPG input is required. Given $\mathbf{x}_{p}$, we compute the condition latent $\mathbf{z}_{p}=E_{\phi}^{p}(\mathbf{x}_{p})$, sample $\mathbf{z}_{0}\sim\mathcal{N}(\mathbf{0},\mathbf{I})$, and solve the learned ODE $d\mathbf{z}_{t}/dt=\mathbf{v}_{\psi}(\mathbf{z}_{t},t,\mathbf{z}_{p})$ from $t=0$ to $t=1$. With an Euler solver using $N$ steps and step size $\Delta t=1/N$, the update is
\begin{equation}
\mathbf{z}_{k+1}
=
\mathbf{z}_{k}
+
\Delta t\,
\mathbf{v}_{\psi}(\mathbf{z}_{k},t_{k},\mathbf{z}_{p}),
\qquad
k=0,\ldots,N-1.
\end{equation}
The final latent $\mathbf{z}_{N}$ is decoded by the frozen ECG decoder to obtain the generated ECG waveform:
\begin{equation}
\hat{\mathbf{x}}_{e}
=
D_{\theta}^{e}(\mathbf{z}_{N}).
\end{equation}

\section{Experiments }
\label{sec:experiments}

\noindent \textbf{Dataset and Preprocessing.}
\label{subsec:dataset}
We evaluate \method{} on MC-MED~\citep{kansal2025multimodal}, a large-scale multimodal clinical monitoring dataset comprising 118,385 adult emergency department visits from 70,545 unique patients at Stanford Health Care between 2020 and 2022. This dataset provides continuous, temporally synchronized physiological waveforms---specifically electrocardiography (ECG) and photoplethysmography (PPG)---alongside visit-level clinical metadata.
MC-MED records PPG at 125~Hz and lead~II ECG at 500~Hz~\citep{fang2025ppgflowecg}. We use preprocessed recordings resampled to 40~Hz for PPG and 120~Hz for ECG. Each synchronized 10-second window thus contains 400 PPG samples ($\mathbf{x}_{p} \in \mathbb{R}^{400 \times 1}$) and 1200 ECG samples ($\mathbf{x}_{e} \in \mathbb{R}^{1200 \times 1}$). Prior to modeling, we apply z-score normalization to each segment. Finally, to ensure signal integrity, we explicitly exclude any windows exhibiting missing values, severe truncation, or invalid signal lengths. We use the official MC-MED random patient-level split, which keeps all visits from each patient in the same partition.

\noindent \textbf{Baselines.}
We compare four baselines: RDDM~\citep{shome2024region} models regions of interest (such as QRS complexes) separately from other regions with diffusion; CardioGAN~\citep{sarkar2021cardiogan} uses an attention-based generator with time- and frequency-domain discriminators; CardioFlow~\citep{nambu2025cardioflow} applies rectified flow to waveforms; and PPGFlowECG~\citep{fang2025ppgflowecg} applies it in a shared PPG--ECG latent space. All models are scored on the same 10-second MC-MED windows at 120~Hz.

\noindent \textbf{Ablation Study.}
To isolate the impact of our core physiological designs, we evaluate \method{} against two ablated variants:
(i) \textit{\method{} w/o Sim}: We remove all simulator guidance during latent flow training. This ablation measures the contribution of simulator guidance during latent flow training.
(ii) \textit{\method{} w/o PAE}: We replace our modality-specific-to-shared PAE architecture with two separate standard autoencoders. This validates the importance of progressively mapping acquisition-specific structures into a shared cardiovascular latent space rather than processing modalities entirely independently.

\subsection{ECG Generation Performance}
\label{subsec:ecg_generation_performance}

\noindent \textbf{Waveform-Level Fidelity.}
MAE and RMSE measure sample-level error; FD and FID compare ECG distributions in waveform space and in a pretrained encoder's feature space, respectively. On MC-MED, \method{} reduces MAE from $0.87$ to $0.71$ relative to PPGFlowECG and also lowers FID and heart-rate MAE (in bpm; Table~\ref{tab:ecg_generation_performance}). Correlation between aligned samples remains modest (Pearson $r=0.1906$), but R-peak F1 for beats matched within 50~ms is $0.74$ for \method{} versus $0.28$ for CardioFlow. Without simulator guidance, F1 drops to $0.62$.

\begin{table}[!ht]
\centering
\caption{ECG generation on MC-MED (mean $\pm$ SD over three runs).}
\label{tab:ecg_generation_performance}
\sisetup{mode=text,reset-text-series=false,round-mode=places,round-precision=2}
\newcommand{\rpeakcell}[3][]{\mbox{#1\num{#2}\ensuremath{{}\pm{}}\num[evaluate-expression=true]{ceil(#3,2)}}}
\footnotesize
\setlength{\tabcolsep}{1.5pt}
\resizebox{\linewidth}{!}{\begin{tabular}{l*{7}{c}}
\toprule
Method & MAE $\downarrow$ & RMSE $\downarrow$ & FD $\downarrow$ & FID $\downarrow$ & HR MAE $\downarrow$ & Pearson $r$ $\uparrow$ & R-peak F1 $\uparrow$ \\
\midrule
RDDM
& $0.95 \pm 0.01$
& $1.41 \pm 0.01$
& $60.34 \pm 0.47$
& $47.97 \pm 0.12$
& $5.48 \pm 0.08$ & $0.0074 \pm 0.0006$ & \rpeakcell{0.2611}{0.0015} \\
CardioGAN
& $1.07 \pm 0.01$
& $1.46 \pm 0.01$
& $74.17 \pm 0.22$
& $47.01 \pm 0.01$
& $4.71 \pm 0.01$ & $-0.0040 \pm 0.0007$ & \rpeakcell{0.0632}{0.0012} \\
CardioFlow
& $0.94 \pm 0.01$
& $1.35 \pm 0.01$
& $58.78 \pm 0.65$
& $86.77 \pm 0.22$
& $6.83 \pm 0.74$ & $0.0277 \pm 0.0004$ & \rpeakcell{0.2766}{0.0013} \\
PPGFlowECG
& $0.87 \pm 0.04$
& $1.32 \pm 0.04$
& $46.64 \pm 1.49$
& $44.17 \pm 0.11$
& $7.08 \pm 0.10$ & $0.0051 \pm 0.0004$ & \rpeakcell{0.1533}{0.0009} \\
\midrule
\rowcolor{methodblue}\textbf{\method{}}
& $\mathbf{0.71 \pm 0.04}$
& $\mathbf{1.07 \pm 0.05}$
& $\mathbf{39.68 \pm 1.96}$
& $\mathbf{37.79 \pm 0.27}$
& $\mathbf{3.94 \pm 0.76}$ & $\mathbf{0.1906 \pm 0.0005}$ & \rpeakcell[\bfseries]{0.7401}{0.0008} \\
\midrule
w/o Sim
& $ 0.84 \pm 0.01$
& $ 1.28 \pm 0.01$
& $ 40.51 \pm 1.20$
& $ 45.90 \pm 0.27 $
& $ 4.94 \pm 0.57 $ & $0.1212 \pm 0.0018$ & \rpeakcell{0.6236}{0.0019} \\
w/o PAE
& $0.85 \pm 0.05$
& $1.26 \pm 0.01$
& $43.10 \pm 0.88$
& $ 44.07 \pm 0.37$
& $ 6.36 \pm 0.07 $ & $0.1471 \pm 0.0022$ & \rpeakcell{0.7094}{0.0024} \\
\bottomrule
\end{tabular}
}
\end{table}

\begin{samepage}
\paragraph{External Dataset Validation.}
Table~\ref{tab:external_waveform} compares \method{} with RDDM, CardioGAN, CardioFlow, and PPGFlowECG on three external datasets. On VitalDB~\citep{lee2022vitaldb}, \method{} has the lowest waveform MAE ($0.61$) and RMSE ($1.30$), while CardioGAN has the lowest heart-rate MAE ($9.17$ bpm). On MIMIC-AFib~\citep{bashar2019noise}, \method{} and PPGFlowECG are nearly tied on MAE and RMSE, but \method{} has the lowest heart-rate MAE ($10.41$ bpm). On BIDMC~\citep{pimentel2017toward}, \method{} has the lowest RMSE ($0.83$) and heart-rate MAE ($3.47$ bpm), while PPGFlowECG has a slightly lower waveform MAE ($0.49$ vs.\ $0.50$).
\par\end{samepage}

\begin{table}[ht]
\centering
\caption{ECG generation on three external datasets (HR MAE in bpm).}
\label{tab:external_waveform}
\footnotesize
\setlength{\tabcolsep}{1pt}
\sisetup{mode=text,reset-text-series=false,round-mode=places,round-precision=2}
\newcommand{\extcell}[3][]{\mbox{{#1\num{#2}}{\scriptsize$\pm$\num[evaluate-expression=true]{ceil(#3,2)}}}}
\begin{tabular*}{\linewidth}{@{\extracolsep{\fill}}l*{9}{c}@{}}
\toprule
& \multicolumn{3}{c}{VitalDB} & \multicolumn{3}{c}{MIMIC-AFib} & \multicolumn{3}{c}{BIDMC} \\
\cmidrule(lr){2-4}\cmidrule(lr){5-7}\cmidrule(lr){8-10}
Method & MAE $\downarrow$ & RMSE $\downarrow$ & HR MAE $\downarrow$
       & MAE $\downarrow$ & RMSE $\downarrow$ & HR MAE $\downarrow$
       & MAE $\downarrow$ & RMSE $\downarrow$ & HR MAE $\downarrow$ \\
\midrule
RDDM
& \extcell{0.7675}{0.0003} & \extcell{1.4144}{0.0003} & \extcell{10.86}{0.08}
& \extcell{0.9969}{0.0005} & \extcell{1.3945}{0.0007} & \extcell{11.97}{0.19}
& \extcell{0.9742}{0.0013} & \extcell{1.3215}{0.0012} & \extcell{3.48}{0.03} \\
CardioGAN
& \extcell{0.8198}{0.00025} & \extcell{1.4109}{0.0004} & \extcell[\bfseries]{9.17}{0.055}
& \extcell{1.0308}{0.0018} & \extcell{1.4130}{0.0013} & \extcell{12.68}{0.145}
& \extcell{1.0245}{0.00095} & \extcell{1.4268}{0.0005} & \extcell{4.35}{0.01} \\
CardioFlow
& \extcell{0.6248}{0.0004} & \extcell{1.3492}{0.0004} & \extcell{16.40}{0.06}
& \extcell{0.5422}{0.0013} & \extcell{0.8663}{0.0025} & \extcell{14.54}{0.12}
& \extcell{0.5186}{0.0010} & \extcell{0.8522}{0.0007} & \extcell{6.83}{0.11} \\
PPGFlowECG
& \extcell{0.6188}{0.0004} & \extcell{1.3074}{0.0004} & \extcell{13.30}{0.09}
& \extcell{0.5219}{0.0027} & \extcell{0.8429}{0.0017} & \extcell{12.81}{0.18}
& \extcell[\bfseries]{0.4945}{0.0014} & \extcell{0.8389}{0.0004} & \extcell{6.54}{0.01} \\
\midrule
\rlap{\smash{\color{methodblue}\rule[-\dp\strutbox]{\linewidth}{\dimexpr\ht\strutbox+\dp\strutbox\relax}}}\textbf{\method{}}
& \extcell[\bfseries]{0.6074}{0.0001} & \extcell[\bfseries]{1.3006}{0.0004} & \extcell{10.49}{0.02}
& \extcell[\bfseries]{0.5178}{0.0009} & \extcell[\bfseries]{0.8377}{0.0009} & \extcell[\bfseries]{10.41}{0.11}
& \extcell{0.4978}{0.0005} & \extcell[\bfseries]{0.8279}{0.0006} & \extcell[\bfseries]{3.47}{0.01} \\
\bottomrule
\end{tabular*}
\end{table}

\paragraph{Morphology and Interval Fidelity.}
On MC-MED, we also assess morphology and intervals beyond global waveform similarity~\citep{weng2026temporalbench}. We use NeuroKit2~\citep{makowski2021neurokit2} to locate fiducial points and compare the seven measurements in Table~\ref{tab:morphology_interval_fidelity} between generated and real ECGs (Appendix~\ref{app:metrics}).
As shown in Table~\ref{tab:morphology_interval_fidelity}, \method{} achieves the best performance across all metrics.
Compared with PPGFlowECG, \method{} reduces PR error from $68.9$ to $59.9$ ms, QRS error from $48.1$ to $38.6$ ms, QT error from $32.5$ to $29.3$ ms, and QTcF error from $35.9$ to $32.2$ ms.
It also improves ST-J60 deviation and both P- and T-wave duration errors, indicating that the generated ECGs better preserve conduction, depolarization, and ST-segment morphology.

\begin{table}[h]
\centering
\caption{Morphology and interval fidelity on MC-MED. We report MAE in milliseconds for intervals and durations, and in normalized amplitude units for ST-J60 deviation.}
\label{tab:morphology_interval_fidelity}
\small
\scalebox{0.87}{
\begin{tabular}{lccccccc}
\toprule
Method
& PR $\downarrow$
& QRS $\downarrow$
& QT $\downarrow$
& QTcF $\downarrow$
& ST-J60 $\downarrow$
& P dur. $\downarrow$
& T dur. $\downarrow$ \\
\midrule
RDDM
& $66.3 \pm 0.4$
& $39.2 \pm 1.1$
& $40.7 \pm 0.6$
& $43.8 \pm 0.6$
& $0.22 \pm 0.0$
& $19.6 \pm 0.2$
& $22.3 \pm 0.1$ \\
CardioGAN
& $65.2 \pm 0.6$
& $43.0 \pm 0.4$
& $56.3 \pm 0.8$
& $61.5 \pm 0.9$
& $0.36 \pm 0.02$
& $27.6 \pm 0.3$
& $24.7 \pm 0.2$ \\
CardioFlow
& $75.8 \pm 1.5$
& $56.8 \pm 0.6$
& $73.9 \pm 1.4$
& $81.1 \pm 1.3$
& $0.20 \pm 0.01$
& $52.8 \pm 0.1$
& $38.9 \pm 0.2$ \\
PPGFlowECG
& $68.9 \pm 0.5$
& $48.1 \pm 0.3$
& $32.5 \pm 0.3$
& $35.9 \pm 0.3$
& $0.16 \pm 0.01$
& $28.8 \pm 0.2$
& $37.7 \pm 0.2$ \\
\midrule
\rowcolor{methodblue}\textbf{\method{}}
& $\mathbf{59.9 \pm 0.9}$
& $\mathbf{38.6 \pm 0.5}$
& $\mathbf{29.3 \pm 0.6}$
& $\mathbf{32.2 \pm 0.5}$
& $\mathbf{0.14 \pm 0.02}$
& $\mathbf{19.3 \pm 0.4 }$
& $\mathbf{21.3 \pm 0.4}$ \\
\bottomrule
\end{tabular}
}
\end{table}

\subsection{Cardiovascular Disease Detection}
\label{subsec:cardiovascular_disease_detection}

To assess whether the generated ECGs preserve clinically useful information, we evaluate downstream multi-label cardiovascular disease detection~\citep{xu2025ecgmoe,xu2025enecg,xie2026enhanced,huang2026lvcg}.
Following prior PPG-to-ECG benchmarks~\citep{fang2025ppgflowecg}, we use visit-level ICD-based cardiovascular labels, including I48, I71, I70, I44, I25, and I50, covering rhythm, conduction, vascular, ischemic, and heart-failure-related conditions.

\begin{table}[h]
\centering
\caption{Quantitative evaluation of multi-label cardiovascular disease classification on the MC-MED dataset. We compare a PPG-based baseline (PaPaGei~\citep{pillai2025papagei}) and \method{}. Higher AUROC indicates better performance.}
\label{tab:mcmed_classification}
\resizebox{0.99\linewidth}{!}{
\begin{tabular}{lccccccc}
\toprule
\multirow{2}{*}{Methods}
& \multicolumn{6}{c}{Disease Categories (AUROC) $\uparrow$}
& \multirow{2}{*}{Macro-AUROC $\uparrow$} \\
\cmidrule(lr){2-7}
& I48 & I71 & I70 & I44 & I25 & I50 & \\
\midrule
PaPaGei
& $0.525 \pm 0.03$
& $0.521 \pm 0.11$
& $0.513 \pm 0.03$
& $0.653 \pm 0.11$
& $0.528 \pm 0.07$
& $0.520 \pm 0.03$
& $0.543 \pm 0.04$ \\
\midrule
\rowcolor{methodblue}\textbf{\method{}}
& $\mathbf{0.697 \pm 0.01}$
& $\mathbf{0.610 \pm 0.02}$
& $\mathbf{0.622 \pm 0.02}$
& $\mathbf{0.656 \pm 0.03}$
& $\mathbf{0.665 \pm 0.01}$
& $\mathbf{0.551 \pm 0.02}$
& $\mathbf{0.633 \pm 0.01}$ \\
\bottomrule
\end{tabular}
}
\end{table}

As shown in Table~\ref{tab:mcmed_classification}, ECGs generated by \method{} achieve higher AUROC than the PPG-based PaPaGei~\citep{pillai2025papagei} baseline across all disease categories, improving Macro-AUROC from $0.543$ to $0.633$.
The gains are particularly clear for I48, I70, and I25, suggesting that the generated ECGs recover clinically relevant electrical patterns that are less accessible from raw PPG alone.
We use the same data split and classifier architecture across all settings.
Since the labels are visit-level ICD codes rather than ECG-only diagnostic annotations, this experiment is intended as a downstream clinical utility evaluation rather than evidence for standalone diagnostic deployment.

\subsection{Qualitative Case Studies}

\begin{figure}[!h]
    \centering
    \includegraphics[width=\linewidth]{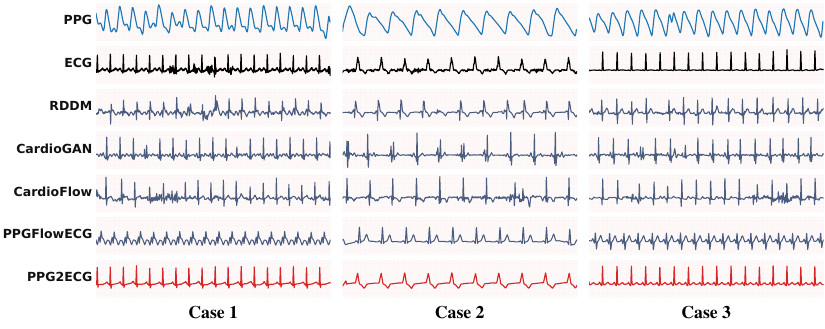}
    \caption{Input PPG, reference ECG, and ECGs generated by four baselines and \method{} for three held-out 10-second MC-MED windows from different patients.}
    \label{fig:case_studies}
\end{figure}

Figure~\ref{fig:case_studies} shows a noisy reference ECG in Case~1 (107~bpm), broad QRS complexes in Case~2 (60~bpm), and a regular rhythm in Case~3 (99~bpm).
\method{} follows the reference beat timing and waveform shape in all three windows, with Pearson correlations of $0.84$, $0.90$, and $0.93$, respectively.
The four baselines generate recognizable beats, but their peaks often miss the reference peaks ($r\leq 0.21$ in these windows).
We z-scored each signal within its window and plotted all ECGs on a common scale in the dataset's per-encounter QRS-upright orientation, without smoothing, time shifts, or further rescaling.

\section{Conclusion}
\label{sec:conclusion}

In this paper, we presented \method{}, a physiology-guided latent rectified flow framework designed to overcome the limitations of purely data-driven PPG-to-ECG generation. Rather than treating this task as a simple statistical cross-modal translation, \method{} explicitly grounds the generative process in the underlying cardiovascular dynamics. We introduced a novel electro-hemodynamic simulator that co-models ECG and PPG via shared cardiac phase dynamics. This simulator provides soft, phase-aware guidance to the Physiology-Aware AutoEncoder to structure the latent space, and explicitly regularizes the latent rectified flow to enforce both ECG-side morphological fidelity and forward ECG-to-PPG hemodynamic consistency. Extensive evaluations on the large-scale MC-MED dataset demonstrate that \method{} consistently outperforms existing baselines, generating ECGs that are not only signal-faithful but also preserve critical clinical intervals and improve downstream cardiovascular disease classification. Ultimately, \method{} effectively bridges the gap between ubiquitous wearable sensing and verifiable clinical cardiac assessment, demonstrating the profound value of integrating mechanistic physiological priors with modern latent generative modeling.

\section*{Acknowledgement}

This research was partially supported by the U.S. National Science Foundation under Award Numbers 2616633, 2442172, 2312502, 2319449, 2211557, 2303037, 2312501, 2425919, 2413417, 2531008, and 2106859, and by the U.S. National Institutes of Health under Award Numbers K25DK135913, RF1NS139325, R01DK143456, U18DP006922, OT2OD038003, R01HL175135, U54OD036472, U54HG012517, and U24DK097771. This research was also partially supported by internal funds and GPU servers provided by the Computer Science Department of Emory University, the SRC JUMP 2.0 Center, Amazon Research Awards, Snapchat Gifts, Optum AI, NEC, the Easton Center, GPU resources from the iTiger GPU cluster~\citep{sharif2025cultivating}, and the U.S. National Science Foundation ACCESS program through allocation CIS260707.

The authors declare no competing interests.

\bibliographystyle{plain}
\bibliography{references}

\appendix

\newpage

\section{The Electro-Hemodynamic Physiological Simulator}
\label{app:simulator_fitting}

\subsection{Euler Discretization}
\label{app:euler_discretization}

We solve the electro-hemodynamic simulator with a fixed-step explicit Euler method.
For any state variable $u(t)$ satisfying $du/dt=v(t)$, the finite-difference approximation is
\begin{equation}
\frac{du}{dt}(t)
\approx
\frac{u(t+\Delta t)-u(t)}{\Delta t},
\end{equation}
which gives the Euler update
\begin{equation}
u(t+\Delta t)=u(t)+v(t)\Delta t.
\end{equation}

For the coupled simulator state $(x(t),y(t),e(t),p(t))$, at time $t_\ell=\ell\Delta t$, the vector field is
\begin{equation}
\mathbf{v}_{\ell}
=
\left(
f_x(x_\ell,y_\ell),
f_y(x_\ell,y_\ell),
f_e(x_\ell,y_\ell,e_\ell,t_\ell;\eta_e),
f_p(x_\ell,y_\ell,p_\ell,t_\ell;\eta_p)
\right).
\end{equation}
The Euler updates are
\begin{align}
x_{\ell+1}
&=
x_{\ell}
+
f_x(x_\ell,y_\ell)\Delta t,\\
y_{\ell+1}
&=
y_{\ell}
+
f_y(x_\ell,y_\ell)\Delta t,\\
e_{\ell+1}
&=
e_{\ell}
+
f_e(x_\ell,y_\ell,e_\ell,t_\ell;\eta_e)\Delta t,\\
p_{\ell+1}
&=
p_{\ell}
+
f_p(x_\ell,y_\ell,p_\ell,t_\ell;\eta_p)\Delta t.
\end{align}

\subsection{Fitting Objective}

\paragraph{Motivation.}
The electro-hemodynamic simulator contains coupled electrical and hemodynamic components: the ECG readout represents upstream cardiac electrical activation, while the PPG readout represents the delayed and smoothed peripheral vascular response. Directly fitting this simulator with a single point-wise waveform loss is often insufficient. A waveform loss may match the global amplitude envelope while missing sharp QRS transitions, systolic upstrokes, or physiologically meaningful event timing. We therefore optimize the simulator parameters with a composite objective that combines waveform-level reconstruction, derivative consistency, and event-level peak alignment. This design encourages the fitted parameters to reproduce both the electrical source dynamics and the downstream hemodynamic response.

Let $e(t)$ and $p(t)$ denote the reference ECG and PPG waveforms, and let $\hat e_\eta(t)$ and $\hat p_\eta(t)$ denote the simulated ECG and PPG waveforms generated by the electro-hemodynamic simulator with parameters $\eta=(\eta_e,\eta_p)$. The total fitting objective is
\begin{equation}
\label{eq:eh_sim_total_loss}
\mathcal{L}_{\mathrm{fit}}(\eta)
=
\lambda_{\mathrm{ecg}}\mathcal{L}_{\mathrm{ecg}}
+
\lambda_{\mathrm{ppg}}\mathcal{L}_{\mathrm{ppg}}
+
\lambda_{\mathrm{deriv}}\mathcal{L}_{\mathrm{deriv}}
+
\lambda_{\mathrm{peak}}\mathcal{L}_{\mathrm{peak}} .
\end{equation}
Unless otherwise specified, we use
$\lambda_{\mathrm{ecg}}=5.0$,
$\lambda_{\mathrm{ppg}}=0.25$,
$\lambda_{\mathrm{deriv}}=3.0$, and
$\lambda_{\mathrm{peak}}=12.0$.

\paragraph{ECG waveform reconstruction.}
The ECG reconstruction term anchors the simulator to the electrical component of the cardiac cycle:
\begin{equation}
\label{eq:sim_ecg_loss}
\mathcal{L}_{\mathrm{ecg}}
=
\frac{1}{T_e}
\sum_{t=1}^{T_e}
\left\|
\hat e_\eta(t)-e(t)
\right\|_2^2 .
\end{equation}
We assign this term a relatively large weight because the ECG is the primary physiological source that drives the downstream hemodynamic response. In practice, this term stabilizes the fitting of QRS-dominated electrical morphology and prevents the simulator from matching peripheral PPG morphology at the expense of unrealistic ECG dynamics.

\paragraph{PPG waveform reconstruction.}
The PPG reconstruction term measures the discrepancy between the simulated and reference peripheral pulse waveforms:
\begin{equation}
\label{eq:sim_ppg_loss}
\mathcal{L}_{\mathrm{ppg}}
=
\frac{1}{T_p}
\sum_{t=1}^{T_p}
\left\|
\hat p_\eta(t)-p(t)
\right\|_2^2 .
\end{equation}
Compared with ECG, PPG is an indirect observation affected by pulse transit time, vascular compliance, peripheral filtering, sensor contact, and amplitude variation. Therefore, we use a smaller weight for this term. This allows PPG to guide the hemodynamic readout without dominating the optimization or distorting the fitted electrical source.

\paragraph{Derivative consistency.}
To preserve local dynamic structure, we further match temporal derivatives:
\begin{equation}
\label{eq:sim_derivative_loss}
\mathcal{L}_{\mathrm{deriv}}
=
\frac{1}{T_e-1}
\sum_{t=1}^{T_e-1}
\left\|
\nabla \hat e_\eta(t)-\nabla e(t)
\right\|_2^2
+
\frac{1}{T_p-1}
\sum_{t=1}^{T_p-1}
\left\|
\nabla \hat p_\eta(t)-\nabla p(t)
\right\|_2^2 ,
\end{equation}
where $\nabla x(t)=x(t+1)-x(t)$ denotes the first-order finite difference. This term emphasizes fast physiological transitions, such as steep QRS slopes in ECG and the systolic upstroke in PPG. It is also less sensitive to slow baseline drift and global amplitude offsets than pure waveform reconstruction. The derivative loss therefore encourages the simulator to reproduce local transition speed and morphology, rather than only matching low-frequency signal shape.

\paragraph{Peak alignment.}
Event timing is central to electro-hemodynamic coupling. We therefore include a peak alignment term that focuses on local neighborhoods around detected physiological events, such as ECG R-peaks and PPG systolic peaks. Let $\mathcal{R}$ denote the set of detected ECG R-peak indices and $\mathcal{S}$ denote the set of detected PPG systolic peak indices. For an event index $\tau$, let $\Omega(\tau)$ denote a local temporal window around the event. In our implementation, the window is defined by $0.20$ seconds before and $0.60$ seconds after the event. The peak alignment loss is
\begin{equation}
\label{eq:sim_peak_loss}
\mathcal{L}_{\mathrm{peak}}
=
\frac{1}{|\mathcal{R}|}
\sum_{\tau\in\mathcal{R}}
\frac{1}{|\Omega_e(\tau)|}
\sum_{t\in\Omega_e(\tau)}
\left\|
\hat e_\eta(t)-e(t)
\right\|_2^2
+
\frac{1}{|\mathcal{S}|}
\sum_{\tau\in\mathcal{S}}
\frac{1}{|\Omega_p(\tau)|}
\sum_{t\in\Omega_p(\tau)}
\left\|
\hat p_\eta(t)-p(t)
\right\|_2^2 .
\end{equation}
This term receives the largest weight because peak locations define key physiological timing events. Matching peaks only through global waveform losses can be unstable: a small temporal shift may cause a large point-wise error even when the morphology is otherwise similar. By explicitly emphasizing local event neighborhoods, the peak loss encourages beat-level synchronization and improves the fitted pulse-arrival relationship between ECG and PPG.

\paragraph{Staged electro-hemodynamic optimization.}
We use a simple curriculum to stabilize fitting. During the initial portion of training, controlled by $\rho_{\mathrm{ecg}}=0.5$, we optimize only ECG-related terms:
\begin{equation}
\label{eq:sim_ecg_warmup}
\mathcal{L}_{\mathrm{warmup}}(\eta)
=
\lambda_{\mathrm{ecg}}\mathcal{L}_{\mathrm{ecg}}
+
\lambda_{\mathrm{deriv}}\mathcal{L}_{\mathrm{deriv}}^{e}
+
\lambda_{\mathrm{peak}}\mathcal{L}_{\mathrm{peak}}^{e},
\end{equation}
where $\mathcal{L}_{\mathrm{deriv}}^{e}$ and $\mathcal{L}_{\mathrm{peak}}^{e}$ denote the ECG components of the derivative and peak losses. After this warm-up stage, we optimize the full electro-hemodynamic objective in Eq.~\ref{eq:eh_sim_total_loss}. This staged design first stabilizes the cardiac electrical source and then introduces the more indirect PPG constraints. It reduces poor local minima in which the simulator prematurely fits delayed peripheral pulse morphology while the underlying ECG dynamics remain inaccurate.

The composite fitting objective assigns different roles to complementary physiological constraints. The ECG waveform loss stabilizes the electrical source, the PPG waveform loss guides the delayed vascular response, the derivative loss preserves fast local dynamics, and the peak alignment loss enforces event-level timing. Together, these terms produce simulator parameters that better capture both ECG morphology and ECG-to-PPG hemodynamic coupling, which are later used as physiological anchors for simulator-guided latent flow training.

\paragraph{ICD-stratified simulator fitting.}
We use the top-100 most frequent ICD-10 codes in the training split to define simulator fitting groups. For each group, representative ECG--PPG beat crops are collected from training samples and used to fit the group-specific simulator parameters $\eta^{\kappa}=(\eta^{\kappa}_{e},\eta^{\kappa}_{p})$ with the composite objective in Eq.~\ref{eq:eh_sim_total_loss}, including the staged warm-up in Eq.~\ref{eq:sim_ecg_warmup}. The fitted parameters are then frozen and used only during training: the fitted pulse-arrival delay $\delta_{\mathrm{PAT}}$ serves as the target of the phase-delay consistency loss $\mathcal{L}_{\mathrm{PAT}}$ in PAE training, and $\eta^{\kappa}$ defines the simulator residual losses $\mathcal{L}^{e}_{\mathrm{sim}}$ and $\mathcal{L}^{p}_{\mathrm{sim}}$ in latent flow training. ICD labels are never used as inputs to the generator and are not required during inference.

\subsection{Consistency Interpretation of Euler Residual Losses}
\label{app:euler_consistency}
The fitted electro-hemodynamic parameters are used to define simulator vector-field residuals on terminal generated beats during latent-flow training. We provide a simple consistency interpretation showing that minimizing these residuals keeps the generated beat close to a trajectory induced by the physiological simulator.
For modality $m\in\{e,p\}$, let $\tilde h^m=(\tilde h^m_1,\ldots,\tilde h^m_{L_m})$ denote a generated beat crop, where $m=e$ corresponds to ECG and $m=p$ corresponds to PPG. Let $f_m$ be the corresponding simulator vector field with fitted parameters $\eta_m$, and let $s^m_\ell=(x^m_\ell,y^m_\ell)$ denote the reference cardiac phase state at step $\ell$. The Euler residual is
\begin{equation}
\label{eq:euler_residual_consistency}
r^m_\ell
=
\frac{\tilde h^m_{\ell+1}-\tilde h^m_\ell}{\Delta t_m}
-
f_m(s^m_\ell,\tilde h^m_\ell,t^m_\ell;\eta_m).
\end{equation}
Assume that $f_m$ is $K_m$-Lipschitz in the waveform state and that the generated beat has bounded average Euler residual,
\begin{equation}
\label{eq:euler_residual_bound_assumption}
\frac{1}{L_m-1}
\sum_{\ell=1}^{L_m-1}
\|r^m_\ell\|_2^2
\leq
\epsilon_m^2 .
\end{equation}
Let $h^{m,\star}$ be the simulator trajectory initialized at the same first sample,
\begin{equation}
h^{m,\star}_{1}=\tilde h^m_{1},
\qquad
h^{m,\star}_{\ell+1}
=
h^{m,\star}_{\ell}
+
\Delta t_m
f_m(s^m_\ell,h^{m,\star}_\ell,t^m_\ell;\eta_m).
\end{equation}
Then the deviation between the generated beat and the simulator-induced trajectory satisfies
\begin{equation}
\label{eq:euler_residual_trajectory_bound}
\max_{\ell\leq L_m}
\|\tilde h^m_\ell-h^{m,\star}_\ell\|_2
\leq
\Delta t_m
\exp(K_m T_m)
\sum_{\ell=1}^{L_m-1}
\|r^m_\ell\|_2,
\end{equation}
where $T_m=(L_m-1)\Delta t_m$ is the beat duration. Consequently, by the Cauchy--Schwarz inequality and Eq.~\ref{eq:euler_residual_bound_assumption},
\begin{equation}
\label{eq:euler_residual_cs}
\sum_{\ell=1}^{L_m-1}\|r^m_\ell\|_2
\leq
\sqrt{L_m-1}
\left(\sum_{\ell=1}^{L_m-1}\|r^m_\ell\|_2^2\right)^{1/2}
\leq
(L_m-1)\,\epsilon_m ,
\end{equation}
so that
\begin{equation}
\label{eq:euler_residual_eps_bound}
\max_{\ell\leq L_m}
\|\tilde h^m_\ell-h^{m,\star}_\ell\|_2
\leq
T_m \exp(K_m T_m)\epsilon_m .
\end{equation}

\paragraph{Proof.}
Let $\delta_\ell=\|\tilde h^m_\ell-h^{m,\star}_\ell\|_2$. By the definition of $r^m_\ell$, the generated beat satisfies $\tilde h^m_{\ell+1}=\tilde h^m_\ell+\Delta t_m f_m(s^m_\ell,\tilde h^m_\ell,t^m_\ell;\eta_m)+\Delta t_m r^m_\ell$. Subtracting the Euler update of $h^{m,\star}$ and applying the triangle inequality,
\begin{equation}
\delta_{\ell+1}
\leq
\delta_\ell
+
\Delta t_m
\left\|
f_m(s^m_\ell,\tilde h^m_\ell,t^m_\ell;\eta_m)
-
f_m(s^m_\ell,h^{m,\star}_\ell,t^m_\ell;\eta_m)
\right\|_2
+
\Delta t_m\|r^m_\ell\|_2 .
\end{equation}
Using the Lipschitz condition gives
\begin{equation}
\delta_{\ell+1}
\leq
(1+K_m\Delta t_m)\delta_\ell
+
\Delta t_m\|r^m_\ell\|_2 .
\end{equation}
Since $\delta_1=0$, unrolling this recursion yields, for every $\ell\leq L_m$,
\begin{equation}
\begin{aligned}
\delta_\ell
&\leq \sum_{j=1}^{\ell-1}
(1+K_m\Delta t_m)^{\ell-1-j}\,\Delta t_m\|r^m_j\|_2 \\
&\leq (1+K_m\Delta t_m)^{L_m-2}\,\Delta t_m
\sum_{j=1}^{L_m-1}\|r^m_j\|_2 \\
&\leq \exp(K_m T_m)\,\Delta t_m
\sum_{j=1}^{L_m-1}\|r^m_j\|_2 ,
\end{aligned}
\end{equation}
where the last step uses $1+u\leq e^{u}$ and $(L_m-2)\Delta t_m\leq T_m$. Taking the maximum over $\ell$ gives Eq.~\ref{eq:euler_residual_trajectory_bound}, and combining it with Eq.~\ref{eq:euler_residual_cs} gives Eq.~\ref{eq:euler_residual_eps_bound}.

This result shows that the simulator-guided Euler residual losses do more than penalize point-wise waveform error. Under mild smoothness assumptions, small residuals imply that the generated ECG or PPG beat remains close to a trajectory generated by the fitted physiological vector field. Therefore, the ECG-side and PPG-side simulator losses constrain generated samples toward electro-hemodynamically plausible dynamics.

\section{Why Physiology-Guided Latent Rectified Flow?}
\label{app:why_lrf}

This section provides a concise theoretical justification for employing latent rectified flow within the \method{} framework. Rather than asserting that rectified flow is universally superior to alternative generative models, our objective is to clarify why it serves as a principled and stable mechanism for PPG-conditioned ECG generation when operating within a structured physiological latent space.

Let $z_p = E_\phi^p(x_p)$ denote the PPG condition latent and $z_e = E_\phi^e(x_e)$ denote the paired ECG latent. Following the optimization of the Physiology-Aware AutoEncoder (PAE), \method{} learns a conditional vector field that transports a Gaussian source latent $z_0 \sim \mathcal{N}(0,I)$ toward the ECG latent distribution, conditioned on $z_p$. For a time variable $t \sim \mathcal{U}(0,1)$, the linear interpolation path is defined as
\[
z_t = (1-t)z_0 + t z_e,
\]
with the target velocity $u_t = z_e-z_0$. The latent rectified-flow objective is then formulated as
\[
\mathcal{L}_{\mathrm{RF}}
=
\mathbb{E}
\left[
\left\|
v_\psi(z_t,t,z_p) - (z_e-z_0)
\right\|_2^2
\right].
\]

\paragraph{Conditional transport interpretation.}
Given a fixed condition $z_p$, the population minimizer of $\mathcal{L}_{\mathrm{RF}}$ is
\[
v^\star(z,t,z_p)
=
\mathbb{E}
\left[
z_e-z_0
\mid
z_t=z,t,z_p
\right].
\]
This follows from the standard result that the conditional mean minimizes the squared-error regression objective, as in rectified flow~\citep{liu2023flow} and flow matching~\citep{lipman2023flow}. Let $p_t(z\mid z_p)$ denote the marginal distribution of $z_t$ conditioned on $z_p$. For any smooth test function $\varphi$, differentiating along the interpolation path yields
\[
\frac{d}{dt}
\mathbb{E}
\left[
\varphi(z_t)
\mid z_p
\right]
=
\mathbb{E}
\left[
\nabla_z \varphi(z_t)^\top (z_e-z_0)
\mid z_p
\right].
\]
Substituting the optimal vector field $v^\star$ results in
\[
\frac{d}{dt}
\mathbb{E}
\left[
\varphi(z_t)
\mid z_p
\right]
=
\mathbb{E}
\left[
\nabla_z \varphi(z_t)^\top v^\star(z_t,t,z_p)
\mid z_p
\right],
\]
which corresponds to the weak form of the continuity equation:
\[
\partial_t p_t(z\mid z_p)
+
\nabla_z \cdot
\left(
p_t(z\mid z_p)v^\star(z,t,z_p)
\right)
=
0.
\]
Consequently, under standard regularity assumptions on the vector field, simulating the ordinary differential equation (ODE) $\frac{d z_t}{dt} = v^\star(z_t,t,z_p)$ deterministically transports the Gaussian source distribution to the conditional ECG latent distribution associated with the observed PPG latent $z_p$. This marginal-preservation property holds for the population minimizer of $\mathcal{L}_{\mathrm{RF}}$ alone. The simulator terms introduced below deliberately bias the learned velocity field toward physiologically consistent terminal states, trading exact marginal matching for physiological plausibility.

\paragraph{Latent-space stability.}
The above derivation formally justifies the generative process in the latent space. To connect this latent transport to target waveform generation, we assume the frozen ECG decoder $D_\theta^e$ is $L_D$-Lipschitz continuous:
\[
\|D_\theta^e(z_1)-D_\theta^e(z_2)\|_2
\leq
L_D\|z_1-z_2\|_2 .
\]
Let $\hat{z}_e$ be the terminal latent produced by the learned flow and let $\hat{x}_e=D_\theta^e(\hat{z}_e)$. The waveform generation error can be bounded as
\[
\|\hat{x}_e-x_e\|_2
\leq
L_D\|\hat{z}_e-z_e\|_2
+
\|D_\theta^e(z_e)-x_e\|_2 .
\]
This inequality shows that the waveform error is upper-bounded by two components: the latent discrepancy $\|\hat{z}_e-z_e\|_2$ and the reconstruction error of the frozen autoencoder. Because PPG-to-ECG generation is a many-to-one inverse problem, the latent discrepancy contains an irreducible part: even a perfect flow samples $\hat{z}_e$ from the conditional distribution $p(z_e\mid z_p)$ rather than recovering the particular paired latent $z_e$, so this term reflects the conditional spread of plausible ECG latents in addition to any transport error. The bound nevertheless motivates the two-stage design: a PAE with small reconstruction error $\|D_\theta^e(z_e)-x_e\|_2$ keeps the second term small, so that the remaining waveform error is governed by the latent-space transport on which the rectified flow and the simulator guidance act.

\paragraph{Role of physiological guidance.}
The simulator-guided objective further regularizes this conditional transport:
\[
\mathcal{L}_{\mathrm{Flow}}
=
\mathcal{L}_{\mathrm{RF}}
+
\lambda_e \mathcal{L}_{\mathrm{sim}}^e
+
\lambda_p \mathcal{L}_{\mathrm{sim}}^p .
\]
In this formulation, $\mathcal{L}_{\mathrm{RF}}$ governs the statistical transport from the Gaussian prior to the empirical ECG latent distribution. Concurrently, the terms $\mathcal{L}_{\mathrm{sim}}^e$ and $\mathcal{L}_{\mathrm{sim}}^p$ bias the predicted terminal state toward trajectories that exhibit valid ECG morphology and forward ECG-to-PPG hemodynamic compatibility. Crucially, because these simulator constraints are applied exclusively during training via gradient signals along the flow trajectory, they do not strictly bound inference to a rigid simulator template. Instead, they function as soft physiological priors, narrowing the set of plausible solutions in the highly underdetermined PPG-to-ECG inverse problem.

In summary, \method{} operates as a regularized conditional transport framework: the PAE constructs a dense, physiology-aware manifold, the rectified flow learns a marginal-preserving deterministic transport upon it, and the simulator guidance anchors the generated distributions to electro-hemodynamically plausible dynamics.

\section{Implementation Details}
\label{Implementation}

\subsection{Training Setup}
\label{app:training_setup}
We use the official MC-MED random patient-level split. We fit the simulator and train all components of \method{} exclusively on visits listed in \texttt{split\_random\_train.csv}. All MC-MED results in the main text and Appendix~\ref{app:additional_results} use windows from visits listed in \texttt{split\_random\_test.csv}. We train in PyTorch on a single NVIDIA H200 GPU with batch size 4. We use PPG resampled to 40~Hz and ECG resampled to 120~Hz. We z-score each modality within each window. After fitting the simulator (Appendix~\ref{app:simulator_fitting}), we train the Physiology-Aware AutoEncoder (PAE). With the PAE frozen, we train the ECG-to-PPG conditional latent rectified flow $G_\rho$ separately on paired training windows. We then train the PPG-to-ECG generator with both the PAE and $G_\rho$ frozen. We apply $\mathcal{L}_{\mathrm{PAT}}$ only to training windows with heart rates of 50--100~bpm, retain windows outside this range for the primary rectified-flow objective, and generate ECGs at inference without estimating PAT.

\paragraph{Baseline evaluation.}
For RDDM, CardioGAN, CardioFlow, and PPGFlowECG, we use the authors' official implementations and released checkpoints with their recommended input preprocessing. We resample inputs to each model's required rate and evaluate a single generated ECG per input on the same 10-second windows at 120~Hz. All step-based methods use 10 sampling steps in the main comparison. For the external evaluation in Table~\ref{tab:external_waveform}, we use \method{} and all four baselines without target-domain fine-tuning or calibration. The released CardioGAN weights were trained in part on BIDMC~\citep{sarkar2021cardiogan}, so its BIDMC result includes prior exposure to that dataset.

\subsection{Physiology-Aware AutoEncoder}
The PAE maps the resampled PPG and ECG segments into a shared electro-hemodynamic latent space. The encoder first processes PPG and ECG with modality-specific stems, then passes the representations through shared blocks with lightweight modality-aware adapters, and finally maps them to a common Gaussian latent posterior. The decoder reconstructs waveforms at their respective input sampling rates using a residual reconstruction branch and a lightweight decomposition branch that models level, trend, and seasonal components. The phase estimators $\Theta_e(\cdot)$ and $\Theta_p(\cdot)$ in $\mathcal{L}_{\mathrm{PAT}}$ are implemented as differentiable soft phase estimators based on local peak-response maps, allowing gradients to propagate through the reconstructed waveforms.
The PAE is optimized with AdamW using a learning rate of $2\times 10^{-5}$ for 120K iterations. The KL weight is set to $5\times 10^{-5}$ with a 5K-iteration linear warmup. To encourage cross-modal latent alignment, we use global posterior alignment with weight $5\times 10^{-5}$ and a 10K-iteration warmup, cross-modal semantic decodability with weight $5\times 10^{-4}$, and bidirectional InfoNCE instance discrimination with weight $10^{-3}$.

\subsection{Latent Rectified Flow}
We freeze the PAE encoder and decoder and train only the PPG-conditioned latent rectified flow. The target is the ECG latent encoded from the paired ECG segment, and the condition is the PPG latent encoded from the synchronized PPG segment. The rectified-flow model uses a Transformer backbone with latent channel size 4, temporal length 50, depth 4, hidden dimension 256, and 4 attention heads. During training, Gaussian noise is linearly interpolated with the target ECG latent, and 4 time points are sampled per forward pass. The model is optimized with the full objective in Eq.~\ref{eq:flow_objective}, which combines the rectified-flow loss with the simulator-guided residual losses.
The latent rectified flow is trained with Adam using a base learning rate of $1\times 10^{-4}$ and betas $(0.9,0.96)$ for 50K iterations. We use warmup and ReduceLROnPlateau scheduling, with 1K warmup steps, patience 1K, and decay factor 0.5. Gradients are clipped to a maximum norm of 1.0. An exponential moving average model is maintained with decay 0.995 and updated every 10 steps.

\subsection{Evaluation Metrics}
\label{app:metrics}
All metrics compare each generated ECG window $\hat{\mathbf{x}}_{e}^{(n)}$ with its paired real ECG window $\mathbf{x}_{e}^{(n)}$, $n=1,\ldots,N_{\mathrm{test}}$, where each window contains $d=1200$ samples (10 s at 120 Hz). The main experiments use per-window z-score normalization (Appendix~\ref{app:training_setup}).

\paragraph{Point-wise fidelity.}
Mean absolute error (MAE) and root mean squared error (RMSE) measure sample-level agreement between paired windows. Both are lower when the reconstructed amplitudes are closer to the reference; RMSE gives larger errors more weight:
\begin{equation}
\label{eq:mae_rmse}
\mathrm{MAE}
=
\frac{1}{N_{\mathrm{test}}\,d}
\sum_{n=1}^{N_{\mathrm{test}}}
\sum_{t=1}^{d}
\left|\hat{x}^{(n)}_{e,t}-x^{(n)}_{e,t}\right|,
\qquad
\mathrm{RMSE}
=
\frac{1}{N_{\mathrm{test}}}
\sum_{n=1}^{N_{\mathrm{test}}}
\sqrt{
\frac{1}{d}
\sum_{t=1}^{d}
\left(\hat{x}^{(n)}_{e,t}-x^{(n)}_{e,t}\right)^{2}
}.
\end{equation}

\paragraph{Distribution-level fidelity.}
FD and FID compare the distribution of generated ECGs with that of real ECGs rather than individual pairs. Given a feature map $\phi$, let $(\boldsymbol{\mu}_{r},\boldsymbol{\Sigma}_{r})$ and $(\boldsymbol{\mu}_{g},\boldsymbol{\Sigma}_{g})$ denote the empirical mean and covariance of $\phi(\mathbf{x}_{e})$ over real windows and of $\phi(\hat{\mathbf{x}}_{e})$ over generated windows. The Fr\'echet distance between the two Gaussian approximations is
\begin{equation}
\label{eq:frechet_distance}
\mathcal{F}_{\phi}
=
\left\|\boldsymbol{\mu}_{r}-\boldsymbol{\mu}_{g}\right\|_{2}^{2}
+
\mathrm{Tr}\left(
\boldsymbol{\Sigma}_{r}
+
\boldsymbol{\Sigma}_{g}
-
2\left(\boldsymbol{\Sigma}_{r}\boldsymbol{\Sigma}_{g}\right)^{1/2}
\right).
\end{equation}
FD evaluates Eq.~\ref{eq:frechet_distance} directly in signal space, with $\phi$ the identity map on the normalized waveform. FID~\citep{heusel2017gans} evaluates Eq.~\ref{eq:frechet_distance} in the embedding space of a pretrained ECG encoder, following SE-Diff~\citep{wang2026se}, and thus measures the distributional discrepancy in a learned ECG feature space.

\paragraph{Heart-rate consistency.}
R-peaks are detected in both the real and generated ECG with NeuroKit2~\citep{makowski2021neurokit2}. For a window whose mean RR interval is $\overline{\mathrm{RR}}$ seconds, the heart rate is $\mathrm{HR}=60/\overline{\mathrm{RR}}$ beats per minute, and
\begin{equation}
\label{eq:hr_mae}
\mathrm{HR\ MAE}
=
\frac{1}{N_{\mathrm{test}}}
\sum_{n=1}^{N_{\mathrm{test}}}
\left|
\mathrm{HR}\big(\hat{\mathbf{x}}_{e}^{(n)}\big)
-
\mathrm{HR}\big(\mathbf{x}_{e}^{(n)}\big)
\right|,
\end{equation}
reported in beats per minute.

\paragraph{Waveform correlation (Pearson $r$).}
We evaluate the last two columns of Table~\ref{tab:ecg_generation_performance} on 10,254 MC-MED test windows without shifting either signal in time. For each generated window $\hat{x}$ and reference window $x$, let $u=\hat{x}-\overline{\hat{x}}$ and $v=x-\bar{x}$ denote the mean-centered waveforms. Pearson $r=\langle u,v\rangle/(\|u\|_2\|v\|_2)$ measures whether amplitudes rise and fall together at corresponding samples. Values range from $-1$ to $1$, with higher values indicating closer agreement in waveform shape.

\paragraph{Beat timing (R-peak F1).}
In the same windows, we match detected R-peaks in the generated ECG to reference R-peaks within 50~ms (six samples at 120~Hz), allowing each peak to be matched at most once. Matched pairs count as true positives (TP), unmatched generated peaks as false positives (FP), and missed reference peaks as false negatives (FN). The score $\mathrm{F1}=2\mathrm{TP}/(2\mathrm{TP}+\mathrm{FP}+\mathrm{FN})$ ranges from 0 to 1, with 1 indicating that every generated and reference peak was matched.

\paragraph{Morphology and interval fidelity.}
Fiducial points are obtained with the discrete-wavelet-transform delineation of NeuroKit2~\citep{makowski2021neurokit2}, which locates the onset and offset of the P wave, the QRS complex, and the T wave around each detected R-peak. From these points, we compute the following measurements for each beat:
\begin{itemize}
    \item \textit{PR interval}: from P-wave onset to QRS onset;
    \item \textit{QRS duration}: from QRS onset to QRS offset (the J point);
    \item \textit{QT interval}: from QRS onset to T-wave offset;
    \item \textit{QTcF}: the Fridericia-corrected QT interval, $\mathrm{QTcF}=\mathrm{QT}/\mathrm{RR}^{1/3}$, with RR in seconds;
    \item \textit{P-wave and T-wave duration}: from onset to offset of the corresponding wave;
    \item \textit{ST-J60 deviation}: the amplitude 60 ms after the J point relative to the isoelectric baseline, in normalized amplitude units.
\end{itemize}
Each measurement is averaged over the beats of a window, and Table~\ref{tab:morphology_interval_fidelity} reports its MAE between generated and real ECGs over test windows. Interval and duration errors are reported in milliseconds.

\section{Additional Results}
\label{app:additional_results}

\subsection{Sampling Steps and Inference Time}
\label{app:sampling_steps}
We vary the number of Euler steps used to generate an ECG from a PPG window on MC-MED (Table~\ref{tab:sampling_steps_latency}). The main evaluation uses 10 steps. Among the four settings tested, five steps give the lowest waveform MAE and RMSE and take 2.47~ms per window, compared with 4.42~ms at ten steps. Heart-rate MAE varies by only 0.2~bpm across the four settings.

\begin{table}[htbp]
\centering
\caption{Generation error and inference time on MC-MED. Time per 10-second ECG is amortized over a batch of 64 windows and includes PPG encoding, flow integration, and ECG decoding.}
\label{tab:sampling_steps_latency}
\small
\begin{tabular}{@{}rcccc@{}}
\toprule
Euler steps & MAE & RMSE & HR MAE (bpm) & Time per ECG (ms) \\
\midrule
5  & 0.66 & 1.02 & 3.9 & 2.47 \\
10 & 0.71 & 1.07 & 4.0 & 4.42 \\
15 & 0.73 & 1.09 & 4.0 & 6.39 \\
20 & 0.74 & 1.10 & 4.1 & 8.40 \\
\bottomrule
\end{tabular}
\end{table}

\subsection{Sensitivity to Signal Normalization}
\label{app:normalization_sensitivity}
The main experiments use per-window z-score normalization. Table~\ref{tab:normalization_sensitivity} compares this choice with min--max normalization on MC-MED. MAE and RMSE depend on the scale of the normalized waveform, and the same change in input scale can affect FD and FID. The timing errors use common units across both settings. Heart-rate MAE differs by 0.01~bpm, and each interval or duration MAE differs by at most 0.7~ms.

\begin{table}[htbp]
\centering
\caption{Effect of normalization on MC-MED. HR MAE is in bpm. Interval and duration errors are MAEs in milliseconds.}
\label{tab:normalization_sensitivity}
\small
\begin{tabular*}{0.95\linewidth}{@{\extracolsep{\fill}}lrrrrr@{}}
\toprule
Normalization & MAE & RMSE & FD & FID & HR MAE \\
\midrule
Min--max & 0.57 & 0.77 & 21.27 & 16.53 & 3.93 \\
Z-score  & 0.71 & 1.07 & 39.68 & 37.79 & 3.94 \\
\bottomrule
\end{tabular*}
\smallskip

\begin{tabular*}{0.95\linewidth}{@{\extracolsep{\fill}}lrrrrrr@{}}
\toprule
Normalization & PR & QRS & QT & QTcF & P duration & T duration \\
\midrule
Min--max & 59.2 & 39.0 & 29.9 & 32.2 & 19.7 & 21.0 \\
Z-score  & 59.9 & 38.6 & 29.3 & 32.2 & 19.3 & 21.3 \\
\bottomrule
\end{tabular*}
\end{table}

\section{Limitations}
We evaluate waveform and heart-rate errors on three external datasets. The morphology, interval, and cardiovascular disease classification results are limited to MC-MED, so their generalization to other clinical settings remains untested.

\section{Broader impacts}
\method{} may have positive impacts for scalable cardiovascular monitoring. Since PPG is widely available in wearable devices and bedside monitors, generating ECG-like signals from PPG could support research on long-term cardiac monitoring, data augmentation, and retrospective analysis when ECG is unavailable or difficult to collect continuously. The proposed physiology-guided framework may also encourage more clinically grounded generative modeling for physiological time series.

\newpage
\section*{NeurIPS Paper Checklist}

\begin{enumerate}

\item {\bf Claims}
    \item[] Question: Do the main claims made in the abstract and introduction accurately reflect the paper's contributions and scope?
    \item[] Answer: \answerYes{}
    \item[] Justification: The main claims in the abstract and introduction match the proposed method and experimental evaluations.
    \item[] Guidelines:
    \begin{itemize}
        \item The answer \answerNA{} means that the abstract and introduction do not include the claims made in the paper.
        \item The abstract and/or introduction should clearly state the claims made, including the contributions made in the paper and important assumptions and limitations. A \answerNo{} or \answerNA{} answer to this question will not be perceived well by the reviewers.
        \item The claims made should match theoretical and experimental results, and reflect how much the results can be expected to generalize to other settings.
        \item It is fine to include aspirational goals as motivation as long as it is clear that these goals are not attained by the paper.
    \end{itemize}

\item {\bf Limitations}
    \item[] Question: Does the paper discuss the limitations of the work performed by the authors?
    \item[] Answer: \answerYes{}
    \item[] Justification: The paper discusses dataset scope, simulator assumptions, signal quality, and clinical deployment limitations.
    \item[] Guidelines:
    \begin{itemize}
        \item The answer \answerNA{} means that the paper has no limitation while the answer \answerNo{} means that the paper has limitations, but those are not discussed in the paper.
        \item The authors are encouraged to create a separate ``Limitations'' section in their paper.
        \item The paper should point out any strong assumptions and how robust the results are to violations of these assumptions (e.g., independence assumptions, noiseless settings, model well-specification, asymptotic approximations only holding locally). The authors should reflect on how these assumptions might be violated in practice and what the implications would be.
        \item The authors should reflect on the scope of the claims made, e.g., if the approach was only tested on a few datasets or with a few runs. In general, empirical results often depend on implicit assumptions, which should be articulated.
        \item The authors should reflect on the factors that influence the performance of the approach. For example, a facial recognition algorithm may perform poorly when image resolution is low or images are taken in low lighting. Or a speech-to-text system might not be used reliably to provide closed captions for online lectures because it fails to handle technical jargon.
        \item The authors should discuss the computational efficiency of the proposed algorithms and how they scale with dataset size.
        \item If applicable, the authors should discuss possible limitations of their approach to address problems of privacy and fairness.
        \item While the authors might fear that complete honesty about limitations might be used by reviewers as grounds for rejection, a worse outcome might be that reviewers discover limitations that aren't acknowledged in the paper. The authors should use their best judgment and recognize that individual actions in favor of transparency play an important role in developing norms that preserve the integrity of the community. Reviewers will be specifically instructed to not penalize honesty concerning limitations.
    \end{itemize}

\item {\bf Theory assumptions and proofs}
    \item[] Question: For each theoretical result, does the paper provide the full set of assumptions and a complete (and correct) proof?
    \item[] Answer: \answerYes{}
    \item[] Justification: The paper provides the assumptions and proof sketch for the Euler-residual consistency result in the appendix.
    \item[] Guidelines:
    \begin{itemize}
        \item The answer \answerNA{} means that the paper does not include theoretical results.
        \item All the theorems, formulas, and proofs in the paper should be numbered and cross-referenced.
        \item All assumptions should be clearly stated or referenced in the statement of any theorems.
        \item The proofs can either appear in the main paper or the supplemental material, but if they appear in the supplemental material, the authors are encouraged to provide a short proof sketch to provide intuition.
        \item Inversely, any informal proof provided in the core of the paper should be complemented by formal proofs provided in appendix or supplemental material.
        \item Theorems and Lemmas that the proof relies upon should be properly referenced.
    \end{itemize}

    \item {\bf Experimental result reproducibility}
    \item[] Question: Does the paper fully disclose all the information needed to reproduce the main experimental results of the paper to the extent that it affects the main claims and/or conclusions of the paper (regardless of whether the code and data are provided or not)?
    \item[] Answer: \answerYes{}
    \item[] Justification: The paper describes the data, preprocessing, model, objectives, metrics, and implementation details.
    \item[] Guidelines:
    \begin{itemize}
        \item The answer \answerNA{} means that the paper does not include experiments.
        \item If the paper includes experiments, a \answerNo{} answer to this question will not be perceived well by the reviewers: Making the paper reproducible is important, regardless of whether the code and data are provided or not.
        \item If the contribution is a dataset and\slash or model, the authors should describe the steps taken to make their results reproducible or verifiable.
        \item Depending on the contribution, reproducibility can be accomplished in various ways. For example, if the contribution is a novel architecture, describing the architecture fully might suffice, or if the contribution is a specific model and empirical evaluation, it may be necessary to either make it possible for others to replicate the model with the same dataset, or provide access to the model. In general. releasing code and data is often one good way to accomplish this, but reproducibility can also be provided via detailed instructions for how to replicate the results, access to a hosted model (e.g., in the case of a large language model), releasing of a model checkpoint, or other means that are appropriate to the research performed.
        \item While NeurIPS does not require releasing code, the conference does require all submissions to provide some reasonable avenue for reproducibility, which may depend on the nature of the contribution. For example
        \begin{enumerate}
            \item If the contribution is primarily a new algorithm, the paper should make it clear how to reproduce that algorithm.
            \item If the contribution is primarily a new model architecture, the paper should describe the architecture clearly and fully.
            \item If the contribution is a new model (e.g., a large language model), then there should either be a way to access this model for reproducing the results or a way to reproduce the model (e.g., with an open-source dataset or instructions for how to construct the dataset).
            \item We recognize that reproducibility may be tricky in some cases, in which case authors are welcome to describe the particular way they provide for reproducibility. In the case of closed-source models, it may be that access to the model is limited in some way (e.g., to registered users), but it should be possible for other researchers to have some path to reproducing or verifying the results.
        \end{enumerate}
    \end{itemize}

\item {\bf Open access to data and code}
    \item[] Question: Does the paper provide open access to the data and code, with sufficient instructions to faithfully reproduce the main experimental results, as described in supplemental material?
    \item[] Answer: \answerYes{}
    \item[] Justification: The clinical dataset (MC-MED) is public dataset, and the current draft provides fully open code.
    \item[] Guidelines:
    \begin{itemize}
        \item The answer \answerNA{} means that paper does not include experiments requiring code.
        \item Please see the NeurIPS code and data submission guidelines (\url{https://neurips.cc/public/guides/CodeSubmissionPolicy}) for more details.
        \item While we encourage the release of code and data, we understand that this might not be possible, so \answerNo{} is an acceptable answer. Papers cannot be rejected simply for not including code, unless this is central to the contribution (e.g., for a new open-source benchmark).
        \item The instructions should contain the exact command and environment needed to run to reproduce the results. See the NeurIPS code and data submission guidelines (\url{https://neurips.cc/public/guides/CodeSubmissionPolicy}) for more details.
        \item The authors should provide instructions on data access and preparation, including how to access the raw data, preprocessed data, intermediate data, and generated data, etc.
        \item The authors should provide scripts to reproduce all experimental results for the new proposed method and baselines. If only a subset of experiments are reproducible, they should state which ones are omitted from the script and why.
        \item At submission time, to preserve anonymity, the authors should release anonymized versions (if applicable).
        \item Providing as much information as possible in supplemental material (appended to the paper) is recommended, but including URLs to data and code is permitted.
    \end{itemize}

\item {\bf Experimental setting/details}
    \item[] Question: Does the paper specify all the training and test details (e.g., data splits, hyperparameters, how they were chosen, type of optimizer) necessary to understand the results?
    \item[] Answer: \answerYes{}
    \item[] Justification: The paper reports preprocessing, splits, hyperparameters, optimizer, model settings, and evaluation metrics.
    \item[] Guidelines:
    \begin{itemize}
        \item The answer \answerNA{} means that the paper does not include experiments.
        \item The experimental setting should be presented in the core of the paper to a level of detail that is necessary to appreciate the results and make sense of them.
        \item The full details can be provided either with the code, in appendix, or as supplemental material.
    \end{itemize}

\item {\bf Experiment statistical significance}
    \item[] Question: Does the paper report error bars suitably and correctly defined or other appropriate information about the statistical significance of the experiments?
    \item[] Answer: \answerYes{}
    \item[] Justification: Table~\ref{tab:ecg_generation_performance} reports across-run standard deviations.
    \item[] Guidelines:
    \begin{itemize}
        \item The answer \answerNA{} means that the paper does not include experiments.
        \item The authors should answer \answerYes{} if the results are accompanied by error bars, confidence intervals, or statistical significance tests, at least for the experiments that support the main claims of the paper.
        \item The factors of variability that the error bars are capturing should be clearly stated (for example, train/test split, initialization, random drawing of some parameter, or overall run with given experimental conditions).
        \item The method for calculating the error bars should be explained (closed form formula, call to a library function, bootstrap, etc.)
        \item The assumptions made should be given (e.g., Normally distributed errors).
        \item It should be clear whether the error bar is the standard deviation or the standard error of the mean.
        \item It is OK to report 1-sigma error bars, but one should state it. The authors should preferably report a 2-sigma error bar than state that they have a 96\% CI, if the hypothesis of Normality of errors is not verified.
        \item For asymmetric distributions, the authors should be careful not to show in tables or figures symmetric error bars that would yield results that are out of range (e.g., negative error rates).
        \item If error bars are reported in tables or plots, the authors should explain in the text how they were calculated and reference the corresponding figures or tables in the text.
    \end{itemize}

\item {\bf Experiments compute resources}
    \item[] Question: For each experiment, does the paper provide sufficient information on the computer resources (type of compute workers, memory, time of execution) needed to reproduce the experiments?
    \item[] Answer: \answerYes{}
    \item[] Justification: The paper reports GPU type, memory, runtime, number of runs, and total compute.
    \item[] Guidelines:
    \begin{itemize}
        \item The answer \answerNA{} means that the paper does not include experiments.
        \item The paper should indicate the type of compute workers CPU or GPU, internal cluster, or cloud provider, including relevant memory and storage.
        \item The paper should provide the amount of compute required for each of the individual experimental runs as well as estimate the total compute.
        \item The paper should disclose whether the full research project required more compute than the experiments reported in the paper (e.g., preliminary or failed experiments that didn't make it into the paper).
    \end{itemize}

\item {\bf Code of ethics}
    \item[] Question: Does the research conducted in the paper conform, in every respect, with the NeurIPS Code of Ethics \url{https://neurips.cc/public/EthicsGuidelines}?
    \item[] Answer: \answerYes{}
    \item[] Justification: The work uses an existing clinical dataset under its access terms and does not attempt patient re-identification.
    \item[] Guidelines:
    \begin{itemize}
        \item The answer \answerNA{} means that the authors have not reviewed the NeurIPS Code of Ethics.
        \item If the authors answer \answerNo, they should explain the special circumstances that require a deviation from the Code of Ethics.
        \item The authors should make sure to preserve anonymity (e.g., if there is a special consideration due to laws or regulations in their jurisdiction).
    \end{itemize}

\item {\bf Broader impacts}
    \item[] Question: Does the paper discuss both potential positive societal impacts and negative societal impacts of the work performed?
    \item[] Answer: \answerYes{}
    \item[] Justification: The paper discusses benefits for wearable monitoring and risks such as misuse, bias, and inappropriate clinical use.
    \item[] Guidelines:
    \begin{itemize}
        \item The answer \answerNA{} means that there is no societal impact of the work performed.
        \item If the authors answer \answerNA{} or \answerNo, they should explain why their work has no societal impact or why the paper does not address societal impact.
        \item Examples of negative societal impacts include potential malicious or unintended uses (e.g., disinformation, generating fake profiles, surveillance), fairness considerations (e.g., deployment of technologies that could make decisions that unfairly impact specific groups), privacy considerations, and security considerations.
        \item The conference expects that many papers will be foundational research and not tied to particular applications, let alone deployments. However, if there is a direct path to any negative applications, the authors should point it out. For example, it is legitimate to point out that an improvement in the quality of generative models could be used to generate Deepfakes for disinformation. On the other hand, it is not needed to point out that a generic algorithm for optimizing neural networks could enable people to train models that generate Deepfakes faster.
        \item The authors should consider possible harms that could arise when the technology is being used as intended and functioning correctly, harms that could arise when the technology is being used as intended but gives incorrect results, and harms following from (intentional or unintentional) misuse of the technology.
        \item If there are negative societal impacts, the authors could also discuss possible mitigation strategies (e.g., gated release of models, providing defenses in addition to attacks, mechanisms for monitoring misuse, mechanisms to monitor how a system learns from feedback over time, improving the efficiency and accessibility of ML).
    \end{itemize}

\item {\bf Safeguards}
    \item[] Question: Does the paper describe safeguards that have been put in place for responsible release of data or models that have a high risk for misuse (e.g., pre-trained language models, image generators, or scraped datasets)?
    \item[] Answer: \answerNA{}
    \item[] Justification: The paper does not release high-risk datasets, large pretrained models, or clinical deployment systems.
    \item[] Guidelines:
    \begin{itemize}
        \item The answer \answerNA{} means that the paper poses no such risks.
        \item Released models that have a high risk for misuse or dual-use should be released with necessary safeguards to allow for controlled use of the model, for example by requiring that users adhere to usage guidelines or restrictions to access the model or implementing safety filters.
        \item Datasets that have been scraped from the Internet could pose safety risks. The authors should describe how they avoided releasing unsafe images.
        \item We recognize that providing effective safeguards is challenging, and many papers do not require this, but we encourage authors to take this into account and make a best faith effort.
    \end{itemize}

\item {\bf Licenses for existing assets}
    \item[] Question: Are the creators or original owners of assets (e.g., code, data, models), used in the paper, properly credited and are the license and terms of use explicitly mentioned and properly respected?
    \item[] Answer: \answerYes{}
    \item[] Justification: The paper cites existing assets and describes their available licenses or access terms.
    \item[] Guidelines:
    \begin{itemize}
        \item The answer \answerNA{} means that the paper does not use existing assets.
        \item The authors should cite the original paper that produced the code package or dataset.
        \item The authors should state which version of the asset is used and, if possible, include a URL.
        \item The name of the license (e.g., CC-BY 4.0) should be included for each asset.
        \item For scraped data from a particular source (e.g., website), the copyright and terms of service of that source should be provided.
        \item If assets are released, the license, copyright information, and terms of use in the package should be provided. For popular datasets, \url{paperswithcode.com/datasets} has curated licenses for some datasets. Their licensing guide can help determine the license of a dataset.
        \item For existing datasets that are re-packaged, both the original license and the license of the derived asset (if it has changed) should be provided.
        \item If this information is not available online, the authors are encouraged to reach out to the asset's creators.
    \end{itemize}

\item {\bf New assets}
    \item[] Question: Are new assets introduced in the paper well documented and is the documentation provided alongside the assets?
    \item[] Answer: \answerYes{}
    \item[] Justification: The released code is documented with setup, preprocessing, training, and evaluation instructions.
    \item[] Guidelines:
    \begin{itemize}
        \item The answer \answerNA{} means that the paper does not release new assets.
        \item Researchers should communicate the details of the dataset\slash code\slash model as part of their submissions via structured templates. This includes details about training, license, limitations, etc.
        \item The paper should discuss whether and how consent was obtained from people whose asset is used.
        \item At submission time, remember to anonymize your assets (if applicable). You can either create an anonymized URL or include an anonymized zip file.
    \end{itemize}

\item {\bf Crowdsourcing and research with human subjects}
    \item[] Question: For crowdsourcing experiments and research with human subjects, does the paper include the full text of instructions given to participants and screenshots, if applicable, as well as details about compensation (if any)?
    \item[] Answer: \answerNA{}
    \item[] Justification: The paper does not involve crowdsourcing or new human-subject experiments.
    \item[] Guidelines:
    \begin{itemize}
        \item The answer \answerNA{} means that the paper does not involve crowdsourcing nor research with human subjects.
        \item Including this information in the supplemental material is fine, but if the main contribution of the paper involves human subjects, then as much detail as possible should be included in the main paper.
        \item According to the NeurIPS Code of Ethics, workers involved in data collection, curation, or other labor should be paid at least the minimum wage in the country of the data collector.
    \end{itemize}

\item {\bf Institutional review board (IRB) approvals or equivalent for research with human subjects}
    \item[] Question: Does the paper describe potential risks incurred by study participants, whether such risks were disclosed to the subjects, and whether Institutional Review Board (IRB) approvals (or an equivalent approval/review based on the requirements of your country or institution) were obtained?
    \item[] Answer: \answerNA{}
    \item[] Justification: The paper does not collect new patient data and uses an existing clinical dataset.
    \item[] Guidelines:
    \begin{itemize}
        \item The answer \answerNA{} means that the paper does not involve crowdsourcing nor research with human subjects.
        \item Depending on the country in which research is conducted, IRB approval (or equivalent) may be required for any human subjects research. If you obtained IRB approval, you should clearly state this in the paper.
        \item We recognize that the procedures for this may vary significantly between institutions and locations, and we expect authors to adhere to the NeurIPS Code of Ethics and the guidelines for their institution.
        \item For initial submissions, do not include any information that would break anonymity (if applicable), such as the institution conducting the review.
    \end{itemize}

\item {\bf Declaration of LLM usage}
    \item[] Question: Does the paper describe the usage of LLMs if it is an important, original, or non-standard component of the core methods in this research? Note that if the LLM is used only for writing, editing, or formatting purposes and does \emph{not} impact the core methodology, scientific rigor, or originality of the research, declaration is not required.
    \item[] Answer: \answerNA{}
    \item[] Justification: LLMs are not used as part of the core method; any use was limited to writing assistance.
    \item[] Guidelines:
    \begin{itemize}
        \item The answer \answerNA{} means that the core method development in this research does not involve LLMs as any important, original, or non-standard components.
        \item Please refer to our LLM policy in the NeurIPS handbook for what should or should not be described.
    \end{itemize}

\end{enumerate}

\end{document}